\documentclass[%
 reprint,
showpacs,preprintnumbers,
 amsmath,amssymb,
 aps,
prb,
]{revtex4-1}

\usepackage{graphicx}
\usepackage{dcolumn}
\usepackage{bm}
\usepackage{color}
\usepackage{ulem}

\allowdisplaybreaks

\begin{document}

\preprint{APS/123-QED}

\title{
Effective bilinear-biquadratic model for noncoplanar ordering in itinerant magnets
}

\author{Satoru Hayami,$^{1}$ Ryo Ozawa,$^2$ and Yukitoshi Motome$^2$}
\affiliation{
$^1$Department of Physics, Hokkaido University, Sapporo 060-0810, Japan\\
$^2$Department of Applied Physics, University of Tokyo, Tokyo 113-8656, Japan
 }

\begin{abstract}
Noncollinear and noncoplanar magnetic textures including skyrmions and vortices act as emergent electromagnetic fields and give rise to novel electronic and transport properties. 
We here report a unified understanding of noncoplanar magnetic orderings emergent from the spin-charge coupling in itinerant magnets. 
The mechanism has its roots in effective multiple spin interactions beyond the conventional Ruderman-Kittel-Kasuya-Yosida (RKKY) mechanism, which are ubiquitously generated in itinerant electron systems with local magnetic moments. 
Carefully examining the higher-order perturbations in terms of the spin-charge coupling, we construct a minimal effective spin model composed of the bilinear and biquadratic interactions with particular wave numbers dictated by the Fermi surface. 
We find that our effective model captures the underlying physics of the instability toward noncoplanar multiple-$Q$ states discovered recently in itinerant magnets: 
a single-$Q$ helical state expected from the RKKY theory is replaced by a double-$Q$ vortex crystal with chirality density waves even for an infinitely small spin-charge coupling on generic lattices [R. Ozawa {\it et al.}, J. Phys. Soc. Jpn. {\bf 85}, 103703 (2016)], and a triple-$Q$ skyrmion crystal with a high topological number of two appears with increasing the spin-charge coupling on a triangular lattice [R. Ozawa, S. Hayami, and Y. Motome, to appear in Phys. Rev. Lett. (arXiv: 1703.03227)]. 
We find that, by introducing an external magnetic field, our effective model exhibits a plethora of multiple-$Q$ states. 
Our findings will serve as a guide for exploring further exotic magnetic orderings in itinerant magnets. 

\end{abstract}
\maketitle

\section{Introduction}
\label{sec: Introduction}
In condensed matter physics, it is a central issue to explore unusual electronic orderings because they bring a major advance in the fundamental physics and open a new path toward applications to next-generation electronics and spintronics devices. 
Among them, magnetic orderings have been intensively investigated from both theory and experiment. 
Various fascinating behaviors have been observed for unusual magnetic structures, such as magnetoelectric effects in noncollinear magnets~\cite{Katsura_PhysRevLett.95.057205,Mostovoy_PhysRevLett.96.067601} and topological Hall effects in noncoplanar magnets~\cite{Loss_PhysRevB.45.13544,Ye_PhysRevLett.83.3737,Ohgushi_PhysRevB.62.R6065}. 
From a theoretical point of view, the important issue is to clarify what kind of interaction is responsible for realizing such unusual magnetic orderings. 
A well-known example is the exchange interaction between localized magnetic moments in Mott insulators. 
Recent studies have revealed that such exchange interactions may result in unusual magnetic states, e.g., a skyrmion crystal with a noncoplanar topological spin texture~\cite{seki2012observation,Adams2012,nagaosa2013topological}, in the presence of the Dzyaloshinskii-Moriya (DM) interaction~\cite{dzyaloshinsky1958thermodynamic,moriya1960anisotropic} coming from the spin-orbit coupling~\cite{rossler2006spontaneous,Yi_PhysRevB.80.054416,Binz_PhysRevLett.96.207202} or frustration between competing interactions~\cite{Okubo_PhysRevLett.108.017206,leonov2015multiply,Lin_PhysRevB.93.064430,Hayami_PhysRevB.93.184413,batista2016frustration}. 

Another representative magnetic interaction has been discussed for peculiar magnetism in rare-earth and other itinerant magnets where localized magnetic moments interact with itinerant electrons. 
The exchange coupling to itinerant electrons gives rise to an effective magnetic interaction between localized moments, which is called the Ruderman-Kittel-Kasuya-Yosida (RKKY) interaction~\cite{Ruderman,Kasuya,Yosida1957}. 
In contrast to the short-ranged interactions in Mott insulators, the RKKY interaction is long-ranged (power-law decay) with changing the sign depending on the distance. 
It often leads to an instability toward helical ordering with a long period structure featured by a single-$Q$ modulation. 

On the other hand, recent theoretical studies have shown that a single-$Q$ helical ordering in itinerant magnets is not stable and taken over by multiple-$Q$ modulated structures~\cite{Martin_PhysRevLett.101.156402,Akagi_JPSJ.79.083711,Hayami_PhysRevB.90.060402,Ozawa_doi:10.7566/JPSJ.85.103703}. 
Such an instability originates from higher-order spin interactions beyond the RKKY interactions, appearing after tracing out the degrees of freedom of itinerant electrons~\cite{Akagi_PhysRevLett.108.096401,Hayami_PhysRevB.90.060402,Ozawa_doi:10.7566/JPSJ.85.103703,Hayami_PhysRevB.94.024424}. 
For example, a triple-$Q$ noncoplanar magnetic order showing the topological Hall effect is realized at particular electron fillings on a triangular lattice, owing to effective four-spin interactions~\cite{Martin_PhysRevLett.101.156402,Akagi_JPSJ.79.083711}. 
Similar mechanism stabilizes double-$Q$ noncoplanar magnetic orders accompanying chirality density waves  at generic filling on generic lattices~\cite{Ozawa_doi:10.7566/JPSJ.85.103703}. 

Above two examples illustrate the importance of effective multiple spin interactions emergent in itinerant magnets in realizing multiple-$Q$ noncoplanar states rather than single-$Q$ helical states. 
The interplay between charge and spin degrees of freedom, however, generates a variety of multiple spin interactions, whose analysis becomes more complicated in the higher-order terms. 
Thus, it is crucial to elucidate essential contributions in itinerant magnets and construct a concise effective model for further exploration of exotic magnetic orderings. 
Such an effective model will be helpful to avoid laborious calculations for the itinerant electron systems, which in general need a tremendous computational cost.

In the present study, we clarify the minimal ingredient to capture the essential physics of exotic magnetism brought by the spin-charge coupling, focusing on the weak coupling regime. 
We show that the origin of noncoplanar magnetic orderings in itinerant magnets is an effective biquadratic interaction specified by particular wave numbers dictated by the Fermi surface. 
We derive an effective spin model with bilinear and biquadratic couplings by examining the dominant contributions in the perturbative expansion in terms of the spin-charge coupling. 
By constructing the phase diagram of the effective spin model on square and triangular lattices by Monte Carlo simulations, we find that our model provides a unified understanding of unconventional multiple-$Q$ magnetic orders previously found in itinerant magnets~\cite{Ozawa_doi:10.7566/JPSJ.85.103703,ozawa2016zero}. 
We confirm the stability of the multiple-$Q$ phases obtained in the spin model by comparing with those in the original itinerant electron model by variational calculations.
We also elucidate the magnetic phase diagram by applying an external magnetic field to our effective model. We find a variety of field-induced multiple-$Q$ phases including different types of double(triple)-$Q$ states on the square (triangular) lattice.

The rest of the paper is organized as follows. 
In Sec.~\ref{sec: Effective Multiple Spin Interactions}, we present the starting itinerant electron model, the Kondo lattice model. 
We discuss how effective multiple spin interactions are generated from the perturbative expansion for the Kondo lattice Hamiltonian with respect to the exchange coupling between itinerant electrons and localized spins. 
By carefully examining contributions from the expansion, we extract a minimal ingredient relevant to exotic magnetic orderings. 
In Sec.~\ref{sec:Effective spin model}, we construct the effective spin model with bilinear-biquadratic interactions defined in momentum space. 
In Sec.~\ref{sec:Multiple-$Q$ magnetic instability}, we discuss the multiple-$Q$ instability in this effective model. 
In Sec.~\ref{sec:Comparison to Itinerant model}, we compare the results between the effective spin model and the original Kondo lattice model by using variational calculations. 
In Sec.~\ref{sec:Effect of magnetic field}, we show the phase diagram under an external magnetic field. 
A plethora of multiple-$Q$ phases is obtained from our Monte Carlo simulations. 
Section~\ref{sec:Discussion} is devoted to a summary and a discussion of candidate materials.

\section{Effective Multiple Spin Interactions in Itinerant magnets}
\label{sec: Effective Multiple Spin Interactions}

In this section, we discuss effective exchange interactions between localized spins based on the perturbative expansion with respect to the exchange coupling in the Kondo lattice model. 
In Sec.~\ref{sec: Model}, we introduce the Hamiltonian of the model. 
In Sec.~\ref{sec: Perturbation expansion}, we present a general expression for the perturbative expansion in the Kondo lattice model. 
Then, in Sec.~\ref{sec: Second-order RKKY interaction}, we discuss the effect of the second-order RKKY interaction, which is not enough to determine magnetic orderings even when the exchange coupling is infinitely small. 
We extract the minimal ingredient to induce noncoplanar magnetic orderings by examining the fourth-order spin interactions in Sec.~\ref{sec: Fourth-order interaction}, and generalize it to higher orders in Sec.~\ref{sec:Higher-order contributions}.

\subsection{Model}
\label{sec: Model}
We begin with a Kondo lattice model consisting of itinerant electrons and localized spins on the square and triangular lattices. 
The Hamiltonian is given by 
\begin{align}
\label{eq:Ham}
\mathcal{H} = -\sum_{i, j,  \sigma} t_{ij} c^{\dagger}_{i\sigma}c_{j \sigma}
+J \sum_{i, \sigma, \sigma'} c^{\dagger}_{i\sigma} \bm{\sigma}_{\sigma \sigma'} c_{i \sigma'}
\cdot \mathbf{S}_i, 
\end{align}
where $c^{\dagger}_{i\sigma}$ ($c_{i \sigma}$) is a creation (annihilation) operator of an itinerant electron at site $i$ and spin $\sigma$. 
The first term represents the kinetic motion of itinerant electrons. 
We consider hopping elements between nearest-neighbor sites, $t_{ij}=t_1$, and third-neighbor sites, $t_{ij}=t_3$, in the following analyses. 
It is noteworthy that qualitative features derived from the model in Eq.~(\ref{eq:Ham}) are expected to hold for other choices of the hopping elements, e.g., second-neighbor hopping instead of $t_3$, whenever the bare magnetic susceptibility shows multiple maxima at symmetry-related wave numbers, as discussed in Sec.~\ref{sec: Second-order RKKY interaction}. Hereafter, we take $t_1=1$ as an energy unit of the model in Eq.~(\ref{eq:Ham}). 
The second term represents the exchange coupling between itinerant electron spins and localized spins. 
$\bm{\sigma}=(\sigma^x,\sigma^y,\sigma^z)$ is the vector of Pauli matrices, $\mathbf{S}_i$ is a localized spin at site $i$ which is regarded as a classical spin with length $|\mathbf{S}_i|=1$, and $J$ is the exchange coupling constant; 
the sign of $J$ is irrelevant for the classical treatment of $\mathbf{S}_i$. 

For the following arguments, it is useful to express the Hamiltonian in Eq.~(\ref{eq:Ham}) in momentum space as 
\begin{align}
\label{eq:Ham_kspace}
\mathcal{H}=\sum_{\mathbf{k},\sigma} \varepsilon_{\mathbf{k}} c^{\dagger}_{\mathbf{k}\sigma}c_{\mathbf{k}\sigma} +
\frac{J}{\sqrt{N}} \sum_{\mathbf{k},\mathbf{q},\sigma, \sigma'} c^{\dagger}_{\mathbf{k}\sigma}\bm{\sigma}_{\sigma \sigma'} c_{\mathbf{k}+\mathbf{q}\sigma'} \cdot \mathbf{S}_{\mathbf{q}}, 
\end{align}
where $c_{\mathbf{k} \sigma}^{\dagger}$ and $c_{\mathbf{k}\sigma}$ are the Fourier transform of $c_{i\sigma}^{\dagger}$ and $c_{i\sigma}$, respectively. 
$\varepsilon_{\mathbf{k}}$ is the energy dispersion of free electrons depending on the lattice structures: 
for the square lattice,  
\begin{align}
\varepsilon_{\mathbf{k}} = -2\sum_{l=1, 2}(t_1 \cos \mathbf{k}\cdot \mathbf{e}_{l} +t_3 \cos  2 \mathbf{k}\cdot \mathbf{e}_{l}), 
\end{align}
where $\mathbf{e}_1=\hat{\mathbf{x}}=(1,0)$ and $\mathbf{e}_2=\hat{\mathbf{y}}=(0,1)$, and 
for the triangular lattice,  
\begin{align}
\varepsilon_{\mathbf{k}} = -2\sum_{l=1,2, 3}(t_1 \cos \mathbf{k}\cdot \mathbf{e}_{l} +t_3 \cos 2 \mathbf{k}\cdot \mathbf{e}_{l}), 
\end{align}
where $\mathbf{e}_1=\hat{\mathbf{x}}$, $\mathbf{e}_2=-\hat{\mathbf{x}}/2+\sqrt{3}\hat{\mathbf{y}}/2$, and $\mathbf{e}_3=-\hat{\mathbf{x}}/2-\sqrt{3}\hat{\mathbf{y}}/2$.  
We set the lattice constant $a=1$ as the length unit. 
In the second term in Eq.~(\ref{eq:Ham_kspace}), $\mathbf{S}_{\mathbf{q}}$ is the Fourier transform of $\mathbf{S}_i$ and $N$ is the number of sites. 
The second term can be regarded as the scattering of itinerant electrons by localized spins with the momentum transfer $\mathbf{q}$.

\subsection{Perturbation expansion}
\label{sec: Perturbation expansion}

\begin{figure}[htb!]
\begin{center}
\includegraphics[width=1.0 \hsize]{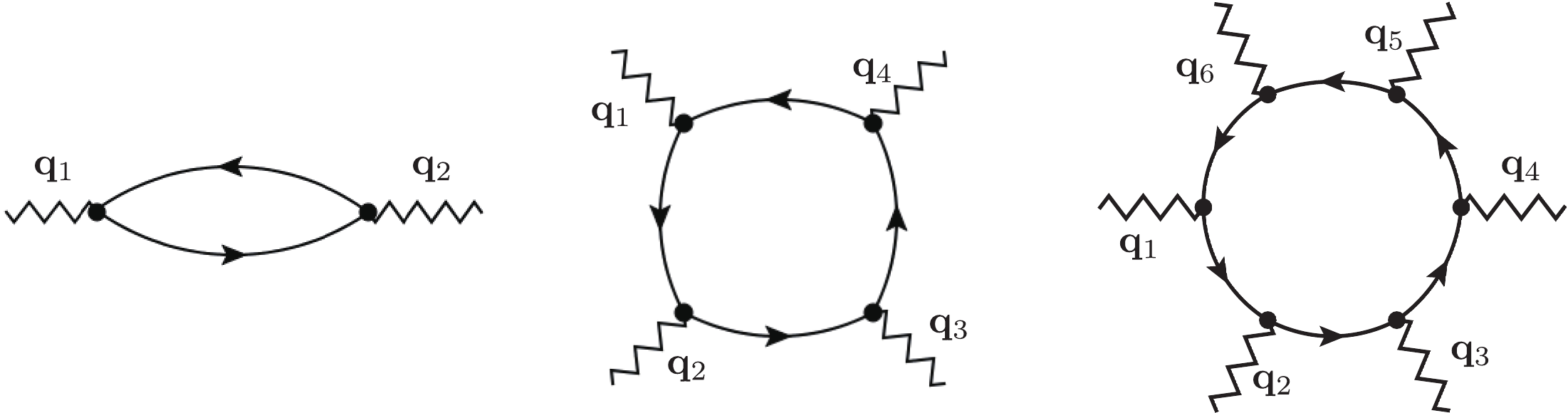} 
\caption{
\label{Fig:diagram}
Feynman diagrams for the first three terms in the perturbative expansion of the free energy in Eq.~(\ref{eq:freeenergy_expand}): $n=1$, $2$, and $3$ in Eq.~(\ref{eq:nth_freeenergy}) from left to right. 
The vertices with wavy lines denote the scattering by localized spins and the solid curves represent the bare propagators of itinerant electrons, $G_{\mathbf{k}}$. 
}
\end{center}
\end{figure}

We consider effective spin interactions mediated by itinerant electrons in the model in Eq.~(\ref{eq:Ham_kspace}). 
Suppose the exchange coupling $J$ is small enough compared to the bandwidth of itinerant electrons, one can expand the free energy of the system with respect to $J$: 
\begin{align}
\label{eq:freeenergy_expand}
F-F^{(0)}&=-T \log 
\left\langle \mathcal{T} \exp  \left( 
-\int^\beta_0 \mathcal{H}' (\tau) d\tau  
\right)
 \right\rangle_{\rm con}   \\
 &=-\frac{T}{2!} \int^{\beta}_0 d \tau_1 \int^{\beta}_0 d \tau_2 
 \langle \mathcal{T} \mathcal{H}' (\tau_1) \mathcal{H}' (\tau_2) \rangle_{\rm con} \nonumber \\
 &\ \ \ - \frac{T}{4!}\int^{\beta}_0 d \tau_1 \cdots  \int^{\beta}_0 d \tau_4 
 \langle \mathcal{T} \mathcal{H}' (\tau_1) \cdots \mathcal{H}' (\tau_4) \rangle_{\rm con} \nonumber \\
&\ \  \ - \cdots \\
\label{eq:freeenergy_expand3}
  &= F^{(2)}+F^{(4)}+\cdots, 
\end{align}
where $\mathcal{H}'$ represents the second term of Eq.~(\ref{eq:Ham_kspace}), $\mathcal{T}$ is the time-ordered product, $\tau$ is the imaginary time, $T$ is the temperature, and $\beta$ is the inverse temperature ($=1/T$; we set the Boltzmann constant as unity). 
$\langle \cdots \rangle_{\rm con}$ stands for the averaged value over the connected Feynman diagrams. 
$F^{(0)}$ represents the free energy from the first term of Eq.~(\ref{eq:Ham_kspace}). 
Note that there is no odd term in the expansion due to the time reversal symmetry of the model. 
The $2n$th-order contribution in the free energy can be expressed in the general form
\begin{widetext} 
\begin{align}
\label{eq:nth_freeenergy}
F^{(2n)}&=\frac{T}{n}\left(\frac{J}{\sqrt{N}}\right)^{2n} 
\sum_{\mathbf{k}, \omega_p}\sum_{\mathbf{q}_1,\cdots,\mathbf{q}_{2n},l} 
G_{\mathbf{k}}G_{\mathbf{k}+\mathbf{q}_1}\cdots G_{\mathbf{k}+\mathbf{q}_1+\cdots+\mathbf{q}_{2n-1}}   
\delta_{\mathbf{q}_1+\mathbf{q}_2+\cdots+\mathbf{q}_{2n},l\mathbf{G}} \sum_{\{P\}}(-1)^{\lambda_P}
\prod_{\nu,\nu'} (\mathbf{S}_{\mathbf{q}_\nu} \cdot \mathbf{S}_{\mathbf{q}_{\nu'}} ), 
\end{align}
\end{widetext}
where $G_{\mathbf{k}} (i \omega_p) = \left[ i \omega_p -(\varepsilon_{\mathbf{k}}-\mu) \right]^{-1}$ is noninteracting Green's function, $\omega_p$ is the Matsubara frequency, $\mu$ is the chemical potential, $\delta$ is the Kronecker delta, and $\mathbf{G}$ is the reciprocal lattice vector ($l$ is an integer). 
The sum of $\{P\}$ is taken for all the combinations of $\nu$ and $\nu'$ (the number of the combinations is $_{2n}\mathrm{C}_{2} \cdot {}_{2n-2}\mathrm{C}_{2} \cdots {}_{2}\mathrm{C}_{2}/(n!)$), and $\lambda_P$ 
is $+1$ ($-1$) for even (odd) permutation.
The product is taken for $1<\nu'<\nu<2n$; 
see also the explicit expressions for the case with $n=1$ and $n=2$ in Eqs.~(\ref{eq:RKKYHam_G}) and (\ref{eq:4thfreeenergy_exact}), respectively. 
In Eq.~(\ref{eq:nth_freeenergy}), we omit the spin dependence of Green's function because the unperturbed Hamiltonian is independent of the spin index. 
The similar expressions were discussed in Ref.~\onlinecite{komarov2016effective}. 

Figure~\ref{Fig:diagram} represents the Feynman diagrams for the first three terms in the expansion of Eq.~(\ref{eq:freeenergy_expand}), namely, $n=1$, $2$, and $3$ in Eq.~(\ref{eq:nth_freeenergy}). 
The diagrams consist of the scattering vertices by localized spins and the bare propagators of itinerant electrons, $G_{\mathbf{k}}$. 
By taking summations in terms of $\mathbf{q}_{\nu}$ ($\nu=1$, $2$, $\cdots$, $2n$) in Eq.~(\ref{eq:nth_freeenergy}), we can obtain the multiple spin interactions at any order of $J$. 
In the following sections, we discuss the specific form of such interactions by focusing on the second order (Sec.~\ref{sec: Second-order RKKY interaction}) and fourth order (Sec.~\ref{sec: Fourth-order interaction}) of $J$, and give the general expression for the higher-order contributions (Sec.~\ref{sec:Higher-order contributions}). 
Hereafter, we do not explicitly indicate the Matsubara frequency dependence in Green's functions for simplicity. 

\subsection{Second-order RKKY interaction}
\label{sec: Second-order RKKY interaction}

First, let us consider the lowest-order contribution in Eq.~(\ref{eq:nth_freeenergy}), i.e., the second order of $J$ ($n=1$). 
This is obtained by substituting $n=1$ into Eq.~(\ref{eq:nth_freeenergy}), which is explicitly written as 
\begin{align}
\label{eq:RKKYHam_G}
F^{(2)}=T
\frac{J^2}{N}  \sum_{\mathbf{k}, \mathbf{q}, \omega_p} G_{\mathbf{k}+\mathbf{q}} G_{\mathbf{k}} \mathbf{S}_{\mathbf{q}}\cdot \mathbf{S}_{-\mathbf{q}}. 
\end{align}
By taking the summation of $\omega_p$, Eq.~(\ref{eq:RKKYHam_G}) turns into 
\begin{align}
\label{eq:RKKYHam}
F^{(2)}&=-J^2 \sum_{\mathbf{q}}\chi_{\mathbf{q}}^0 \mathbf{S}_{\mathbf{q}}\cdot \mathbf{S}_{-\mathbf{q}}, 
\end{align}
where $\chi_{\mathbf{q}}^0$ is the bare susceptibility of itinerant electrons, 
\begin{align}
\chi_{\mathbf{q}}^0 &=\frac{T}{N}\sum_{\mathbf{k},\omega_p} G_{\mathbf{k}+\mathbf{q}}G_{\mathbf{k}}  \\
\label{eq:chi0}
&= \frac{1}{N} \sum_{\mathbf{k}} \frac{f(\varepsilon_{\mathbf{k}})-f(\varepsilon_{\mathbf{k}+\mathbf{q}})}{\varepsilon_{\mathbf{k}+\mathbf{q}}-\varepsilon_{\mathbf{k}}}. 
\end{align}
Eq.~(\ref{eq:RKKYHam}) gives a pairwise interaction between localized spins, which is called the RKKY interaction~\cite{Ruderman,Kasuya,Yosida1957}. 
The sign and amplitude of the interaction depend on the band structure and electron filling through Eq.~(\ref{eq:chi0}). 

\begin{figure}[htb!]
\begin{center}
\includegraphics[width=0.5 \hsize]{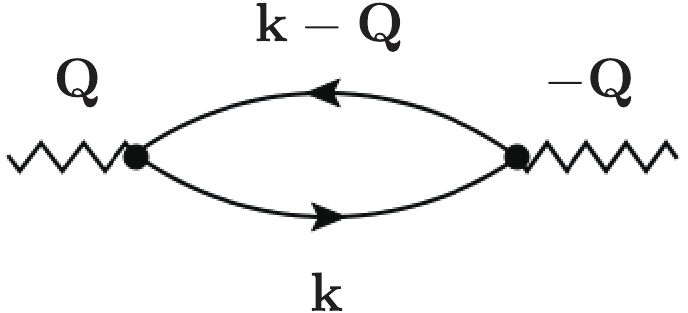} 
\caption{
\label{Fig:diagram_2nd}
Feynman diagram for the lowest second-order contribution in the single-$Q$ helical state with ordering vector $\mathbf{Q}$ [Eq.~(\ref{eq:F(2)_helical})]. 
}
\end{center}
\end{figure}

The magnetic state that optimizes the RKKY interaction in Eq.~(\ref{eq:RKKYHam}) is a single-$Q$ helical (spiral) state, whose spin structure is represented by 
\begin{align}
\label{eq:spin_helical}
\mathbf{S}_i=(\cos \mathbf{Q}\cdot \mathbf{r}_i,\sin \mathbf{Q}\cdot \mathbf{r}_i,0).
\end{align}
Here, $\mathbf{Q}$ is the ordering vector defining the pitch and direction of the spiral, which is dictated by the peak of $\chi_{\mathbf{q}}^0$ in Eq.~(\ref{eq:chi0}). Note that the spiral axis in the helical state in Eq.~(\ref{eq:spin_helical}) is arbitrary because of the spin rotational symmetry of the RKKY interaction in Eq.~(\ref{eq:RKKYHam}). 
The reason why the helical state is preferred is understood from the normalization condition $|\mathbf{S}_i|=1$, which is equivalent to $\sum_{\mathbf{q}}|\mathbf{S}_{\mathbf{q}}|^2=N$: 
the helical state, in which 
$|\mathbf{S}_{\mathbf{Q}}|^2=|\mathbf{S}_{-\mathbf{Q}}|^2=N/2$ and $\mathbf{S}_{\mathbf{q}}=0$ for $\mathbf{q}\neq \pm\mathbf{Q}$, gives the lowest energy of Eq.~(\ref{eq:RKKYHam}).
In other words, any superposition with another wave number or any higher harmonics leads to an energy cost compared to the helical state. 
Thus, the second-order free energy for the helical state is given by
\begin{align}
\label{eq:F(2)_helical}
F^{(2)}&=-J^2 (\chi_{\mathbf{Q}}^0 |S_{\mathbf{Q}}|^2+\chi_{-\mathbf{Q}}^0 |S_{-\mathbf{Q}}|^2).  
\end{align} 
The corresponding Feynman diagram is shown in Fig.~\ref{Fig:diagram_2nd}. 

An important point in the RKKY analysis is that there are still four(six)fold degeneracy in the helical state on the square (triangular) lattice. Among them, the twofold degeneracy comes from the chiral symmetry in the centrosymmetric Bravais lattice structures, while the remaining two(three)fold degeneracy from the $C_4$ ($C_6$) rotational symmetry of the square (triangular) lattice structures. 
The latter degeneracy plays a role in realizing multiple-$Q$ orderings as discussed in Sec.~\ref{sec: Fourth-order interaction}. 

\begin{figure}[ht!]
\begin{center}
\includegraphics[width=1.0 \hsize]{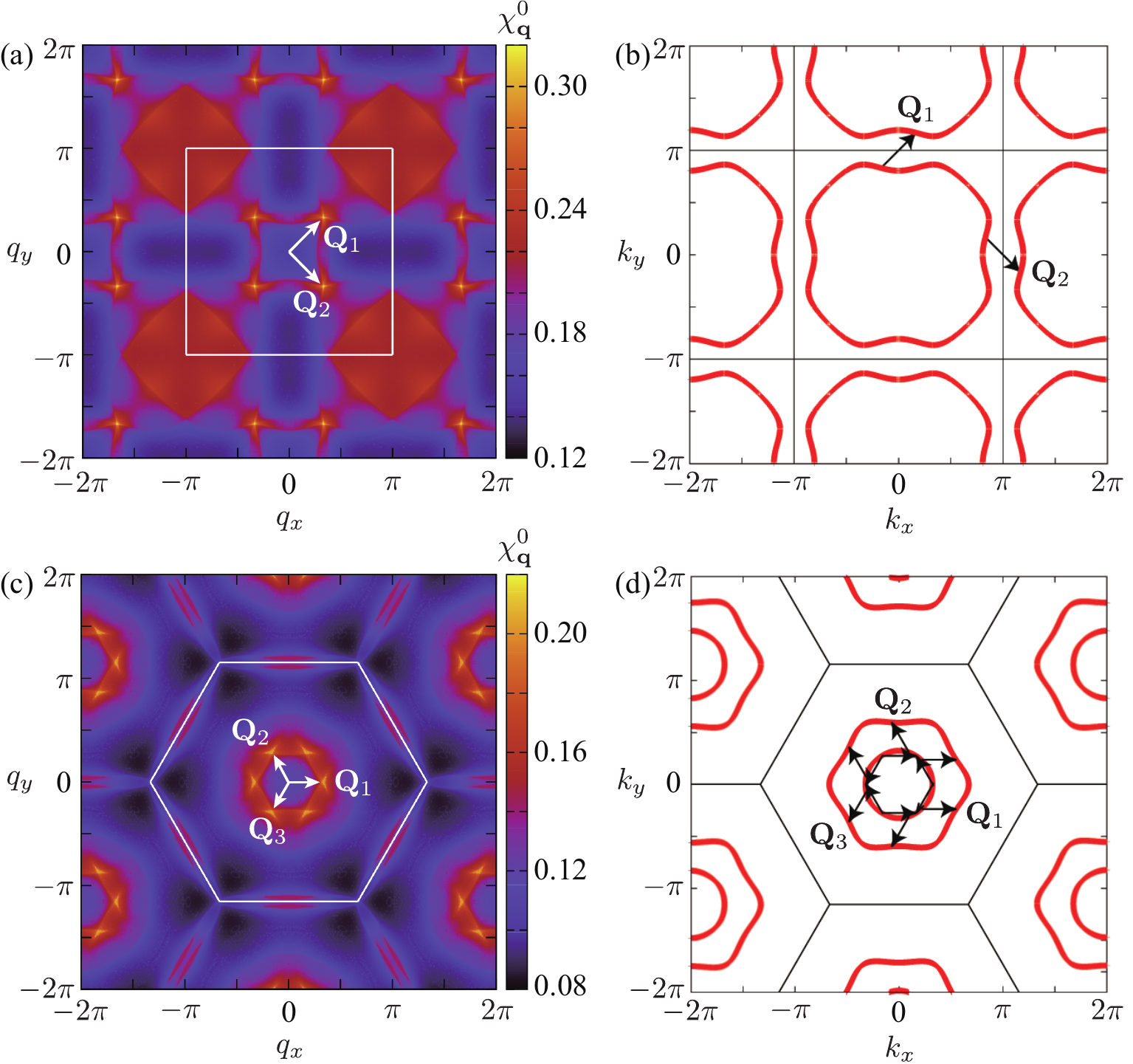} 
\caption{
\label{Fig:suscep}
(a), (c) The contour plots of the bare susceptibility $\chi_{\mathbf{q}}^0$ for (a) the square lattice model at $t_3=-0.5$ and $\mu=0.98$ and (c) the triangular lattice model at $t_3=-0.85$ and $\mu=-3.5$. 
$\chi_{\mathbf{q}}^0$ exhibits maxima at $\mathbf{Q}_1$ and $\mathbf{Q}_2$ in (a), while $\mathbf{Q}_1$, $\mathbf{Q}_2$, and $\mathbf{Q}_3$ in (c). In both cases, $\mathbf{Q}_\nu$ are connected with each other by the rotational symmetry operation of the lattice structure. 
The squares and hexagons in the figures represent the first Brillouin zone. 
(b), (d) The Fermi surfaces corresponding to (a) and (c), respectively. 
$\mathbf{Q}_\nu$ are the vectors giving the maxima of $\chi^0_{\mathbf{q}}$ in (a) and (c). 
}
\end{center}
\end{figure}

Let us describe how the degeneracy arises from the rotational symmetry by showing the momentum dependence of the bare susceptibility. $\chi_{\mathbf{q}}^0$ possesses multiple peaks reflecting the rotational symmetry of the lattice structure. 
This is demonstrated in Fig.~\ref{Fig:suscep}: 
Figs.~\ref{Fig:suscep}(a) and~\ref{Fig:suscep}(c) show $\chi_{\mathbf{q}}^0$ on the square lattice with $t_3=-0.5$ and $\mu=0.98$ and the triangular lattice with $t_3=-0.85$ and $\mu=-3.5$, respectively. 
The corresponding Fermi surfaces are also shown in Figs.~\ref{Fig:suscep}(b) and \ref{Fig:suscep}(d). 
The bare susceptibility shows multiple peaks at the wave numbers for which the Fermi surface is nested, and the wave numbers respect the rotational symmetry of the system. 
The former square lattice case has four peak structures at $\mathbf{Q}_1=\pm(2\pi/6, 2\pi/6)$ and $\mathbf{Q}_2=R(\pi/2)\mathbf{Q}_1$, and the latter triangular lattice case shows six peak structures at $\mathbf{Q}_1=(2\pi/6,0)$, $\mathbf{Q}_2=R(2\pi/3)\mathbf{Q}_1$ and $\mathbf{Q}_3=R(4\pi/3)\mathbf{Q}_1$; 
here, $R(\theta)$ represents the rotational operator by $\theta$. 
As described above, at the level of the RKKY interaction in Eq.~(\ref{eq:RKKYHam}), the single-$Q$ helical order is realized by choosing the ordering vector $\mathbf{Q}$ out of these multiple peaks. 

In this way, the second-order free energy, i.e., the RKKY interaction, favors the helical ordering with the single-$Q$ modulation represented by Eq.~(\ref{eq:spin_helical}). 
However, the ground state in the Kondo lattice model is not given by the helical state but by noncoplanar multiple-$Q$ states even for the $J \to 0$ limit. 
The striking result was originally found at particular electron fillings where the Fermi surface has perfect nesting~\cite{Martin_PhysRevLett.101.156402} or multiple connections~\cite{Akagi_JPSJ.79.083711,Akagi_PhysRevLett.108.096401,Hayami_PhysRevB.90.060402}, while recently generalized to generic fillings where the Fermi surface has no special property except for the rotational symmetry~\cite{Ozawa_doi:10.7566/JPSJ.85.103703}. 
The fundamental mechanism is that the system tends to lift the degeneracy due to the rotational symmetry of the lattice structure through the higher-order contributions of the free energy, as described in the following sections.

\subsection{Fourth-order interaction}
\label{sec: Fourth-order interaction}

\begin{figure}[htb!]
\begin{center}
\includegraphics[width=1.0 \hsize]{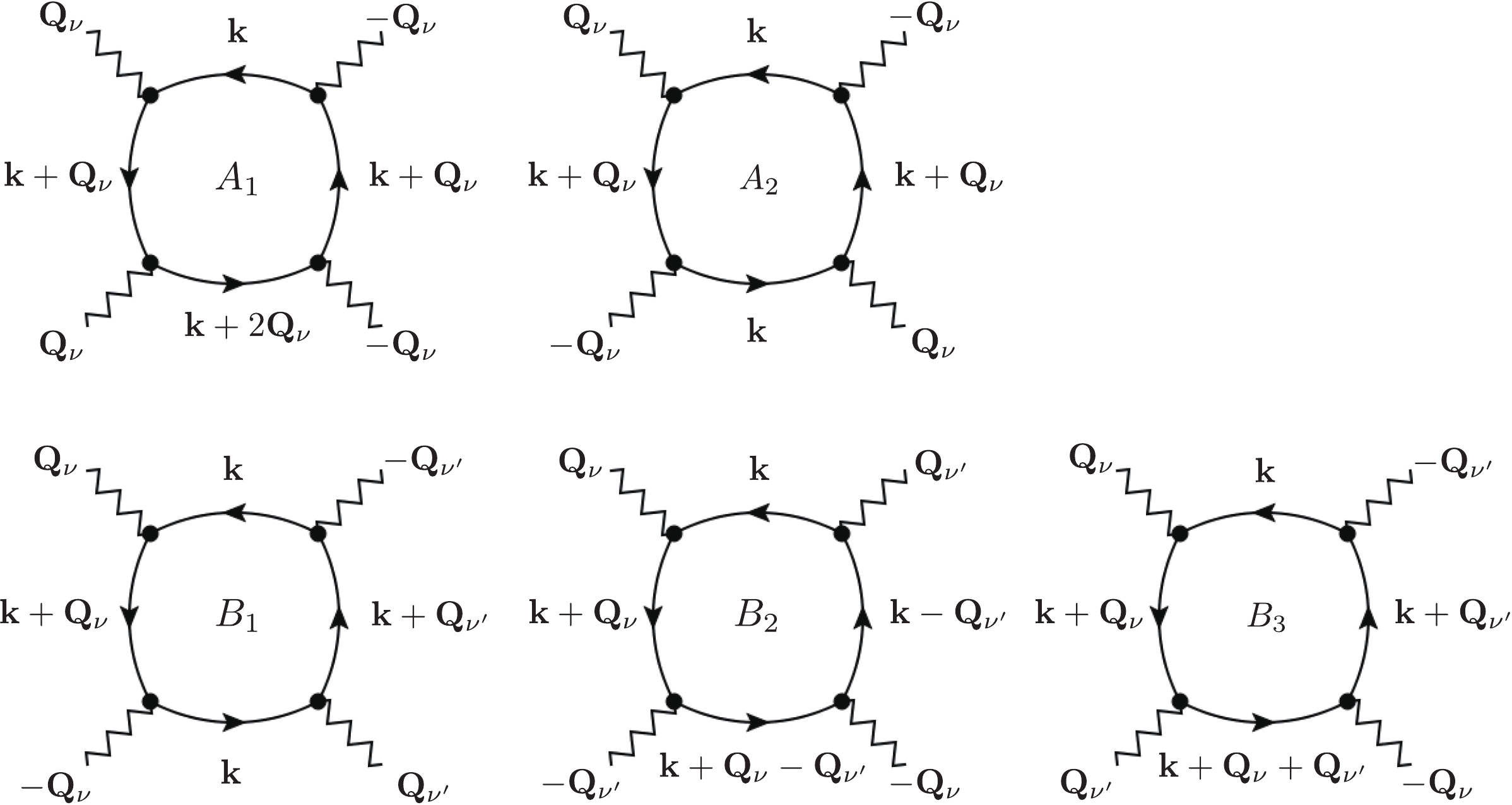} 
\caption{
\label{Fig:diagram_4th}
Feynman diagrams for the coefficients $A_1$, $A_2$, $B_1$, $B_2$, and $B_3$ in the fourth-order contributions. See Eqs.~(\ref{eq:F4})-(\ref{eq:B3}).
}
\end{center}
\end{figure}

Next, we consider the fourth-order contribution of the free energy, Eq.~(\ref{eq:nth_freeenergy}) with $n=2$. 
It is explicitly written as 
\begin{widetext}
\begin{align}
\label{eq:4thfreeenergy_exact}
F^{(4)}=\frac{T}{2}\frac{J^4}{N^2} \sum_{\mathbf{k}, \omega_p}\sum_{\mathbf{q}_1,\mathbf{q}_2,\mathbf{q}_3,\mathbf{q}_4,l} &
G_{\mathbf{k}}G_{\mathbf{k}+\mathbf{q}_1}G_{\mathbf{k}+\mathbf{q}_1+\mathbf{q}_2}G_{\mathbf{k}+\mathbf{q}_1+\mathbf{q}_2+\mathbf{q}_3} \delta_{\mathbf{q}_1+\mathbf{q}_2+\mathbf{q}_3+\mathbf{q}_4,l\mathbf{G}} 
\nonumber \\
&\times \left[ 
(\mathbf{S}_{\mathbf{q}_1}\cdot \mathbf{S}_{\mathbf{q}_2})
(\mathbf{S}_{\mathbf{q}_3}\cdot \mathbf{S}_{\mathbf{q}_4})
+
(\mathbf{S}_{\mathbf{q}_1}\cdot \mathbf{S}_{\mathbf{q}_4})
(\mathbf{S}_{\mathbf{q}_2}\cdot \mathbf{S}_{\mathbf{q}_3})
-
(\mathbf{S}_{\mathbf{q}_1}\cdot \mathbf{S}_{\mathbf{q}_3})
(\mathbf{S}_{\mathbf{q}_2}\cdot \mathbf{S}_{\mathbf{q}_4})
\right].  
\end{align}
\end{widetext}
This gives the four-spin interactions, which play an important role in lifting the degeneracy between the helical orderings and leads to the instability toward multiple-$Q$ orderings. 
For discussing such degeneracy lifting, it is enough to take into account the wave numbers for the multiple maxima in the bare susceptibility: $\mathbf{q}=\pm\mathbf{Q}_1$ and $\pm\mathbf{Q}_2$ ($\pm\mathbf{Q}_1$, $\pm\mathbf{Q}_2$, and $\pm\mathbf{Q}_3$) for the square (triangular) lattice. 
In the following, we consider the scattering processes satisfying $\mathbf{q}_1+\mathbf{q}_2+\mathbf{q}_3+\mathbf{q}_4 =0 $, i.e., $l=0$ in Eq.~(\ref{eq:4thfreeenergy_exact}); the special cases with $\mathbf{q}_1+\mathbf{q}_2+\mathbf{q}_3+\mathbf{q}_4 = \mathbf{G}$ were discussed for $2\mathbf{Q}_{\nu} = \mathbf{G}$~\cite{Martin_PhysRevLett.101.156402,Akagi_JPSJ.79.083711,Akagi_PhysRevLett.108.096401,Hayami_PhysRevB.90.060402,hayami_PhysRevB.91.075104} and $4\mathbf{Q}_{\nu} = \mathbf{G}$~\cite{Hayami_PhysRevB.94.024424} ($\nu=1,2,3$). 
Then, the fourth-order free energy is given by the sum of five types of multiple spin interactions: 
\begin{widetext}
\begin{align}
\label{eq:F4}
F^{(4)}&=F^{(4)}_1+F^{(4)}_2+F^{(4)}_3+F^{(4)}_4+F^{(4)}_5, \\
\label{eq:F41}
F^{(4)}_1&=\frac{J^4}{N} \sum_{\nu}
(2A_1- A_2)
(\mathbf{S}_{\mathbf{Q}_\nu}\cdot \mathbf{S}_{\mathbf{Q}_\nu})
(\mathbf{S}_{-\mathbf{Q}_\nu}\cdot \mathbf{S}_{-\mathbf{Q}_\nu}),  \\
\label{eq:F42}
F^{(4)}_2&=\frac{J^4}{N} \sum_{\nu}
(2A_2)
(\mathbf{S}_{\mathbf{Q}_\nu}\cdot \mathbf{S}_{-\mathbf{Q}_\nu})^2,  
  \\
  \label{eq:F43}
F^{(4)}_3&=4\frac{J^4}{N} \sum_{\nu, \nu'}
(B_1+ B_2 -  B_3)
(\mathbf{S}_{\mathbf{Q}_\nu}\cdot \mathbf{S}_{-\mathbf{Q}_\nu})
(\mathbf{S}_{\mathbf{Q}_{\nu'}}\cdot \mathbf{S}_{-\mathbf{Q}_{\nu'}}), \\
\label{eq:F44}
F^{(4)}_4&=4\frac{J^4}{N} \sum_{\nu, \nu'}
(-B_1+ B_2 +   B_3)
(\mathbf{S}_{\mathbf{Q}_\nu}\cdot \mathbf{S}_{\mathbf{Q}_{\nu'}})
(\mathbf{S}_{-\mathbf{Q}_\nu}\cdot \mathbf{S}_{-\mathbf{Q}_{\nu'}}), \\
\label{eq:F45}
F^{(4)}_5&=4\frac{J^4}{N} \sum_{\nu, \nu'}
(B_1- B_2 + B_3)
(\mathbf{S}_{\mathbf{Q}_\nu}\cdot \mathbf{S}_{-\mathbf{Q}_{\nu'}})
(\mathbf{S}_{-\mathbf{Q}_{\nu'}}\cdot \mathbf{S}_{\mathbf{Q}_\nu}), 
\end{align}
\end{widetext}
where the sums in Eqs.~(\ref{eq:F43})-(\ref{eq:F45}) are taken for $\nu>\nu'$. The coefficients are given by 
\begin{align}
\label{eq:A1}
A_1 &= \frac{T}{N}\sum_{\mathbf{k}, \omega_p}(G_{\mathbf{k}})^2 G_{\mathbf{k}-\mathbf{Q}_\nu}G_{\mathbf{k}+\mathbf{Q}_\nu}, \\
\label{eq:A2}
A_2 &= \frac{T}{N}\sum_{\mathbf{k}, \omega_p}(G_{\mathbf{k}})^2 (G_{\mathbf{k}+\mathbf{Q}_\nu})^2, \\
\label{eq:B1}
B_1 &= \frac{T}{N}\sum_{\mathbf{k}, \omega_p}(G_{\mathbf{k}})^2 G_{\mathbf{k}+\mathbf{Q}_\nu}G_{\mathbf{k}+\mathbf{Q}_{\nu'}}, \\
\label{eq:B2}
B_2 &= \frac{T}{N}\sum_{\mathbf{k}, \omega_p}(G_{\mathbf{k}})^2 G_{\mathbf{k}+\mathbf{Q}_\nu}G_{\mathbf{k}-\mathbf{Q}_{\nu'}}, \\
\label{eq:B3}
B_3 &= \frac{T}{N}\sum_{\mathbf{k}, \omega_p} G_{\mathbf{k}} G_{\mathbf{k}+\mathbf{Q}_\nu}G_{\mathbf{k}+\mathbf{Q}_{\nu'}}G_{\mathbf{k}+\mathbf{Q}_\nu+\mathbf{Q}_{\nu'}}.   
\end{align}
Each contribution is expressed by the diagram in Fig.~\ref{Fig:diagram_4th}. 
$A_1$ and $A_2$ represent the scattering processes by a single wave number, while $B_1$, $B_2$, and $B_3$ by multiple wave numbers.

\begin{figure}[htb!]
\begin{center}
\includegraphics[width=1.0 \hsize]{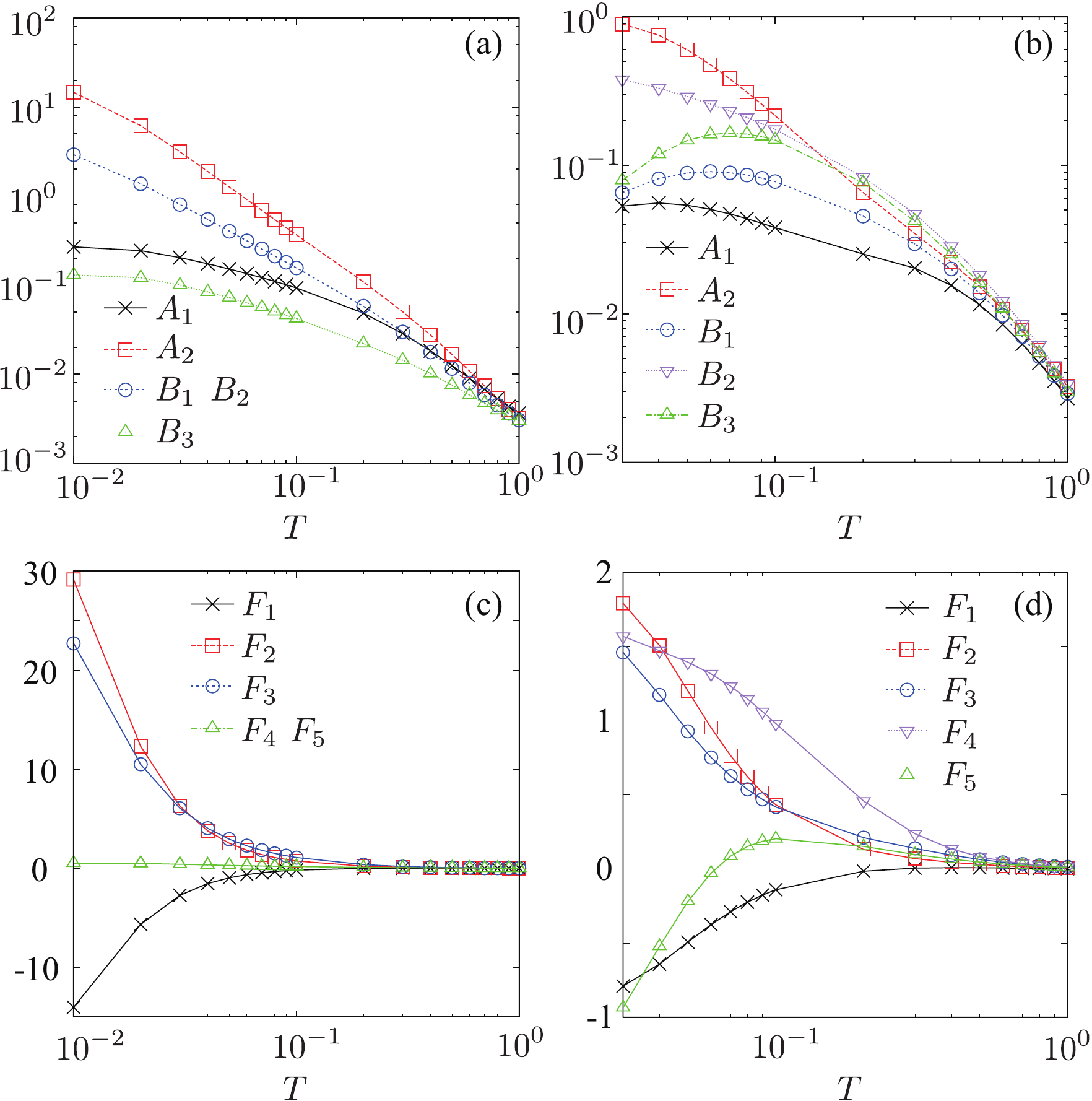} 
\caption{
\label{Fig:4th}
(a), (b) Temperature dependences of the coefficients in Eqs.~(\ref{eq:A1})-(\ref{eq:B3}) at (a) $t_3=-0.5$ and $\mu=0.98$ on the square lattice and (b) $t_3=-0.85$ and $\mu=-3.5$ on the triangular lattice. 
The data are calculated for the system size $N=3600^2$ and the number of Matsubara frequency 8000. 
(c), (d) Temperature dependences of fourth-order free energy in Eqs.~(\ref{eq:F41})-(\ref{eq:F45}) for 
the same parameters in (a) and (b), respectively. 
}
\end{center}
\end{figure}

Let us discuss which term plays a dominant role among the five types of multiple spin interactions in Eqs.~(\ref{eq:F41})-(\ref{eq:F45}). 
Figures~\ref{Fig:4th}(a) and \ref{Fig:4th}(b) compares the coefficients $A_1$, $A_2$, $B_1$, $B_2$, and $B_3$ for two sets of parameters used in Fig.~\ref{Fig:suscep}. 
In both cases, the coefficient $A_2$ becomes dominant in the low-temperature limit. 
This indicates that $F_2^{(4)}$ is the most important contribution among the fourth-order multiple spin interactions, as confirmed in Figs.~\ref{Fig:4th}(c) and \ref{Fig:4th}(d).

It is worth noting that $F_2^{(4)}$ in Eq.~(\ref{eq:F42}) is in the form of the biquadratic interaction with a positive coefficient $A_2$. 
The positive biquadratic interaction, in general, favors a noncollinear spin configuration. 
For example, a triple-$Q$ noncoplanar ordering is realized on a triangular lattice, when the positive biquadratic interaction is enhanced by Fermi surface connections~\cite{Akagi_PhysRevLett.108.096401,Hayami_PhysRevB.90.060402}. 

\subsection{Higher-order contributions}
\label{sec:Higher-order contributions}

\begin{figure}[htb!]
\begin{center}
\includegraphics[width=1.0 \hsize]{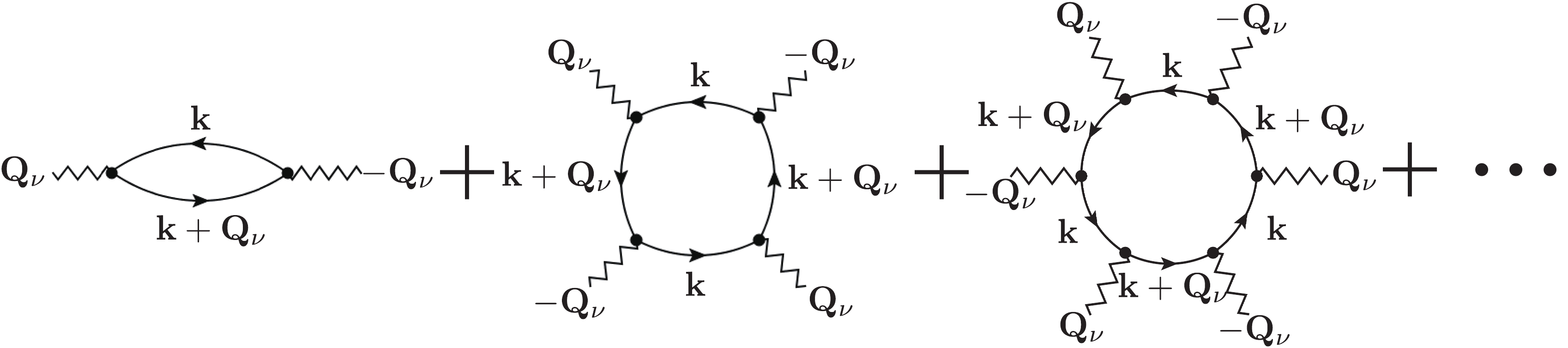} 
\caption{
\label{Fig:diagram_rpa}
The Feynman diagrams describing the dominant contributions among all the spin scattering processes. 
}
\end{center}
\end{figure}

The higher-order contributions can be straightforwardly expressed by the Feynman diagrams similar to those in Fig.~\ref{Fig:diagram_4th}. 
By analyzing the contributions from each diagram, we find that the scattering process proportional to $(\mathbf{S}_{\mathbf{Q}_\nu} \cdot \mathbf{S}_{-\mathbf{Q}_\nu})^n$ gives the most important contribution in each order. 
Note that the dominant $F_2^{(4)}$ in Eq.~(\ref{eq:F42}) is the case with $n=2$. 
Thus, the general form of the dominant contribution in the free energy at the $2n$th order is given by 
\begin{align}
\label{eq:F2nessential}
F^{(2n)}_{(\mathbf{Q},-\mathbf{Q})} = &\frac{2^n T}{n} \left(\frac{J}{\sqrt{N}}\right)^{2n} 
\nonumber \\   &\times
  \sum_{\mathbf{k}, \omega_p,\nu}
(G_{\mathbf{k}})^n (G_{\mathbf{k}+\mathbf{Q}_\nu})^n 
(\mathbf{S}_{\mathbf{Q}_\nu} \cdot \mathbf{S}_{-\mathbf{Q}_\nu})^n. 
\end{align}
The sum of the dominant contributions up to the infinite orders can be summarized in a compact form 
\begin{widetext}
\begin{align}
\label{eq:freeenergy_QQ}
 F_{(\mathbf{Q},-\mathbf{Q})} &= F^{(2)}_{(\mathbf{Q},-\mathbf{Q})}+F^{(4)}_{(\mathbf{Q},-\mathbf{Q})}+ \cdots + F^{(2n)}_{(\mathbf{Q},-\mathbf{Q})} + \cdots \\
\label{eq:renofreeenergy}
&= 2T \frac{J^2}{N} \sum_{\mathbf{k},\omega_p, \nu} \left[ 
\frac{G_{\mathbf{k}} G_{\mathbf{k}+\mathbf{Q}_{\nu}}}{1-2 (J^2/N) G_{\mathbf{k}} G_{\mathbf{k}+\mathbf{Q}_{\nu}}(\mathbf{S}_{\mathbf{Q}_\nu} \cdot \mathbf{S}_{-\mathbf{Q}_\nu}) }
\right](\mathbf{S}_{\mathbf{Q}_\nu} \cdot \mathbf{S}_{-\mathbf{Q}_\nu}).  
\end{align}
\end{widetext}
The corresponding Feynman diagrams are shown in Fig.~\ref{Fig:diagram_rpa}. 

We find that the dominant terms in Eq.~(\ref{eq:freeenergy_QQ}) contribute in a different way depending on the order of the expansion: 
the $(4l+2)$th-order terms proportional to $TG^{2l+1}_{\mathbf{k}}G^{2l+1}_{\mathbf{k}+\mathbf{Q}}<0$ tend to favor a single-$Q$ coplanar order, while the $4l$th-order ones proportional to $TG^{2l}_{\mathbf{k}}G^{2l}_{\mathbf{k}+\mathbf{Q}}>0$ tend to favor multiple-$Q$ noncoplanar order. 
This implies that Eq.~(\ref{eq:freeenergy_QQ}) is divided into two groups, which are phenomenologically represented by a bilinear interaction $(\mathbf{S}_{\mathbf{Q}_\nu}\cdot \mathbf{S}_{-\mathbf{Q}_{\nu}})$ and a biquadratic interaction $(\mathbf{S}_{\mathbf{Q}_\nu}\cdot \mathbf{S}_{-\mathbf{Q}_{\nu}})^2$. 
On the basis of this observation, we construct an effective model in Sec.~\ref{sec:Effective spin model}. 

We note that the free energy in Eq.~(\ref{eq:renofreeenergy}) obtained the partial summation of the dominant contributions converges in the limit of zero temperature, while each term is divergent, as inferred in Figs.~\ref{Fig:4th}(c) and \ref{Fig:4th}(d). 
We discuss the effect of the higher-order contributions by evaluating Eq.~(\ref{eq:renofreeenergy}) in Sec.~\ref{sec:Comparison to Itinerant model}.

\section{Effective spin model}
\label{sec:Effective spin model}
\subsection{Bilinear-biquadratic model in momentum space}

The perturbation expansion in Sec.~\ref{sec: Effective Multiple Spin Interactions} indicates that many different types of effective spin interactions can contribute to the magnetic ordering in itinerant magnets. 
The careful comparison between different terms, however, gives an insight into the dominant contribution, as discussed in Secs.~\ref{sec: Fourth-order interaction} and \ref{sec:Higher-order contributions}. 
Based on the observations, we propose an effective spin model in the weak coupling regime by including the contributions from the bilinear and biquadratic interactions. 
The Hamiltonian is given by 
\begin{align}
\label{eq:effHam_spin}
\mathcal{H}=  2\sum_\nu
\left[ -\tilde{J}\mathbf{S}_{\mathbf{Q_{\nu}}} \cdot \mathbf{S}_{-\mathbf{Q_{\nu}}}
+\tilde{K} (\mathbf{S}_{\mathbf{Q_{\nu}}} \cdot \mathbf{S}_{-\mathbf{Q_{\nu}}})^2 \right], 
\end{align}
where the sum is taken for $\nu=1,2$ ($1,2,3$) for the square (triangular) lattice, and $\mathbf{Q}_\nu$ are the wave numbers for the multiple peaks of the bare susceptibility $\chi_\mathbf{q}^0$\; $\tilde{J}$ and $\tilde{K}$ are the coupling constants for bilinear and biquadratic interactions, respectively, in momentum space. 
We take $\tilde{J}>0$ and $\tilde{K}>0$, following the most dominant contributions to each term,  Eqs.~(\ref{eq:F(2)_helical}) and (\ref{eq:F42}). 
Hereafter, we set $\tilde{J}=1$ and change the parameter $K = N\tilde{K}$ in the following calculations. 

The model in Eq.~(\ref{eq:effHam_spin}) is a bilinear-biquadratic model defined in momentum space. 
The bilinear-biquadratic model has been studied in real space: the interactions are defined between the localized spins at particular sites, like $\mathbf{S}_i \cdot \mathbf{S}_j$ and $(\mathbf{S}_i \cdot \mathbf{S}_j)^2$ in Refs.~\onlinecite{Sivardiere_PhysRevB.5.1126,Chen_PhysRevB.7.4284,Harada_PhysRevB.65.052403}. 
In our model, however, the interactions are defined for the Fourier components of spins whose wave numbers are dictated by the Fermi surface~\footnote{The effective model can be applicable to the case of $2\mathbf{Q}_{\nu} = \mathbf{G}$ and $4\mathbf{Q}_{\nu} = \mathbf{G}$ ($\nu=1,2,3$)}. 

The perturbative argument in Sec.~\ref{sec: Effective Multiple Spin Interactions} suggests that the bilinear $\tilde{J}$ term is dominant over the biquadratic $\tilde{K}$ term, as the former originates from the RKKY interaction proportional to $J^2$ [Eq.~(\ref{eq:F(2)_helical})], while the most dominant contribution to the latter is proportional to $J^4$ [Eq.~(\ref{eq:F42})] in the weak coupling limit. 
Nevertheless, we will study the effective model in Eq.~(\ref{eq:effHam_spin}) up to $K \sim \tilde{J}$ in the following sections, since the dominant contributions from the higher-order perturbation are phenomenologically renormalized into the effective bilinear and biquadratic interactions, as discussed in Sec.~\ref{sec:Higher-order contributions}. 
We will demonstrate that the extension of the model to the nonperturbative region $K \sim \tilde J$ is indeed useful to discuss the magnetic instabilities appearing in the original Kondo lattice model in Eq.~(\ref{eq:Ham}).

\subsection{Monte Carlo simulation}

In order to obtain the magnetic phase diagram of the effective spin model in Eq.~(\ref{eq:effHam_spin}), we perform classical Monte Carlo simulations at low temperatures. 
Our simulations are carried out with the standard Metropolis local updates. 
In the following, we present the results for the systems with $N=96^2$ sites under periodic boundary conditions in both the square and triangular lattice cases; we confirm that the system size dependence is small by performing the simulations for $N=48^2$ and $72^2$. 
In each simulation, we perform simulated annealing to find the low-energy configuration in the following way. 
We gradually reduce the temperature with a rate $T_{n+1}=\alpha T_{n}$ where $T_n$ is the temperature in the $n$th step. We set the initial temperature $T_0=0.1$-$1.0$ and take the coefficient of geometrical cooling $\alpha=0.9995$-$0.9999$. 
The final temperature, which is typically taken at $T=0.01$, is reached by spending totally $10^5$-$10^6$ Monte Carlo sweeps. 
At the final temperature, we perform $10^5$-$10^6$ Monte Carlo sweeps for measurements after $10^5$-$10^6$ steps for thermalization. 

We calculate the structure factors for spin and scalar chirality to identify each magnetic phase. 
The spin structure factor $S_s(\mathbf{q})$ is given by 
\begin{align}
\label{eq:spinstructurefactor}
S_s(\mathbf{q})= S_s^{xx}(\mathbf{q})+S_s^{yy}(\mathbf{q})+S_s^{zz}(\mathbf{q}),  
\end{align}
where
\begin{align}
S_s^{\alpha \alpha}(\mathbf{q})&= \frac{1}{N} \sum_{j,l} \langle S_j^{\alpha} S_l^{\alpha} \rangle e^{i \mathbf{q}\cdot (\mathbf{r}_j-\mathbf{r}_l)},
\end{align}
with $\alpha=x,y,z$. 
We also compute 
\begin{align}
S_s^{\perp}(\mathbf{q})= S_s^{xx}(\mathbf{q})+S_s^{yy}(\mathbf{q}), 
\end{align}
in the presence of the external magnetic field applied along the $z$ direction. 
We also introduce the following notation for the magnetic moments at $\mathbf{q}$ components: 
\begin{align}
\label{eq:m_q}
&m_{\mathbf{q}}=\sqrt{S_s(\mathbf{q})/N}. 
\end{align}

In addition, we calculate the chirality structure factor. 
For the square lattice, it is defined by 
\begin{align}
\label{eq:chiralstructurefactor_squ}
S_{\chi^{\rm sc}}(\mathbf{q})= \frac{1}{N}\sum_{i,j} \langle \chi^{\rm sc}_{i}
\chi^{\rm sc}_{j}\rangle e^{i \mathbf{q}\cdot (\mathbf{r}_i-\mathbf{r}_j)}, 
\end{align}
where the local scalar chirality at site $i$ is introduced as 
$\chi^{\rm sc}_i = \mathbf{S}_{i} \cdot (\mathbf{S}_{i+\hat{x}}\times \mathbf{S}_{i+\hat{y}})
+\mathbf{S}_{i} \cdot (\mathbf{S}_{i-\hat{x}}\times \mathbf{S}_{i-\hat{y}})
-\mathbf{S}_{i} \cdot (\mathbf{S}_{i-\hat{x}}\times \mathbf{S}_{i+\hat{y}})
-\mathbf{S}_{i} \cdot (\mathbf{S}_{i+\hat{x}}\times \mathbf{S}_{i-\hat{y}})$. 
Meanwhile, the chirality structure factor on the triangular lattice is defined separately for the upward and downward triangles as 
\begin{align}
\label{eq:chiralstructurefactor}
S^{\mu}_{\chi^{\rm sc}}(\mathbf{q})= \frac{1}{N}\sum_{\mathbf{R},\mathbf{R}' \in \mu} \langle \chi^{\rm sc}_{\mathbf{R}}
\chi^{\rm sc}_{\mathbf{R}'}\rangle e^{i \mathbf{q}\cdot (\mathbf{R}-\mathbf{R}')}, 
\end{align}
where $\mathbf{R}$ and $\mathbf{R}'$ represent the position vectors at the centers of triangles, 
and $\mu=(u, d)$ represent upward and downward triangles, respectively. 
Note that connecting the center of gravity in the upward and downward triangles leads to the honeycomb networks. 
In Eq.~(\ref{eq:chiralstructurefactor}), $\chi^{\rm sc}_{\mathbf{R}}= \mathbf{S}_j \cdot (\mathbf{S}_k \times \mathbf{S}_l)$, where $j,k,l$ are the sites on the triangle at $\mathbf{R}$ in the counterclockwise order. 
We also introduce the scalar chirality at $\mathbf{q}$ components: 
\begin{align}
\label{eq:chi_q}
&\chi^{\rm sc}_{\mathbf{q}}=\sqrt{S^u_{\chi^{\rm sc}}(\mathbf{q})/N+S^d_{\chi^{\rm sc}}(\mathbf{q})/N}. 
\end{align}

\section{Multiple-$Q$ magnetic instability}
\label{sec:Multiple-$Q$ magnetic instability}
In this section, we investigate magnetic instabilities appearing in the bilinear-biquadratic model in Eq.~(\ref{eq:effHam_spin}), which is derived as an effective model for the Kondo lattice model with small $J/t$. 
In Sec.~\ref{sec:Magnetic phase diagram}, we elucidate the phase diagrams on the square and triangular lattices, in which the helical state expected from the RKKY interaction is no longer stable and replaced by multiple-$Q$ states. 
In Sec.~\ref{sec:Energy comparison}, we discuss why the multiple-$Q$ states have lower energy than the single-$Q$ helical state. 

\subsection{Phase diagram}
\label{sec:Magnetic phase diagram}

\subsubsection{Square lattice}
\label{sec:Square lattice}

\begin{figure}[htb!]
\begin{center}
\includegraphics[width=0.7 \hsize]{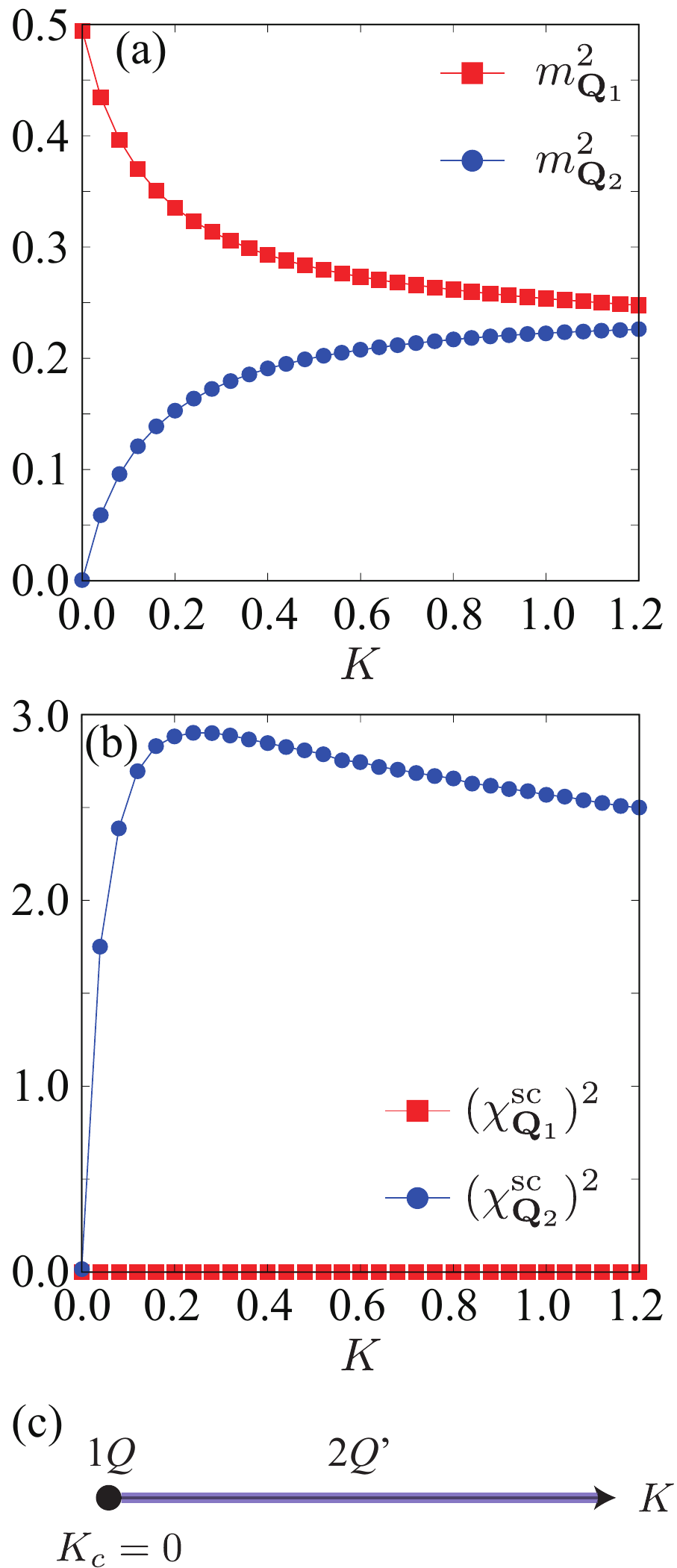} 
\caption{
\label{Fig:Kdep_phase_squ}
(a), (b) 
$K$ dependences of the $\mathbf{Q}_\nu$ components of (a) the squared magnetization and (b) the squared scalar chirality obtained by Monte Carlo simulations for the model in Eq.~(\ref{eq:effHam_spin}) on the square lattice.
The parameters are $\mathbf{Q}_1=(2\pi/6,2\pi/6)$ and $\mathbf{Q}_2=R(\pi/2)\mathbf{Q}_1$. 
(c) Phase diagram for the square lattice case. 
The single-$Q$ helical state ($1Q$) is limited at $K=0$, and the double-$Q'$ chiral stripe ($2Q'$) is realized for all $K>0$. 
}
\end{center}
\end{figure}

\begin{figure*}[htb!]
\begin{center}
\includegraphics[width=0.75 \hsize]{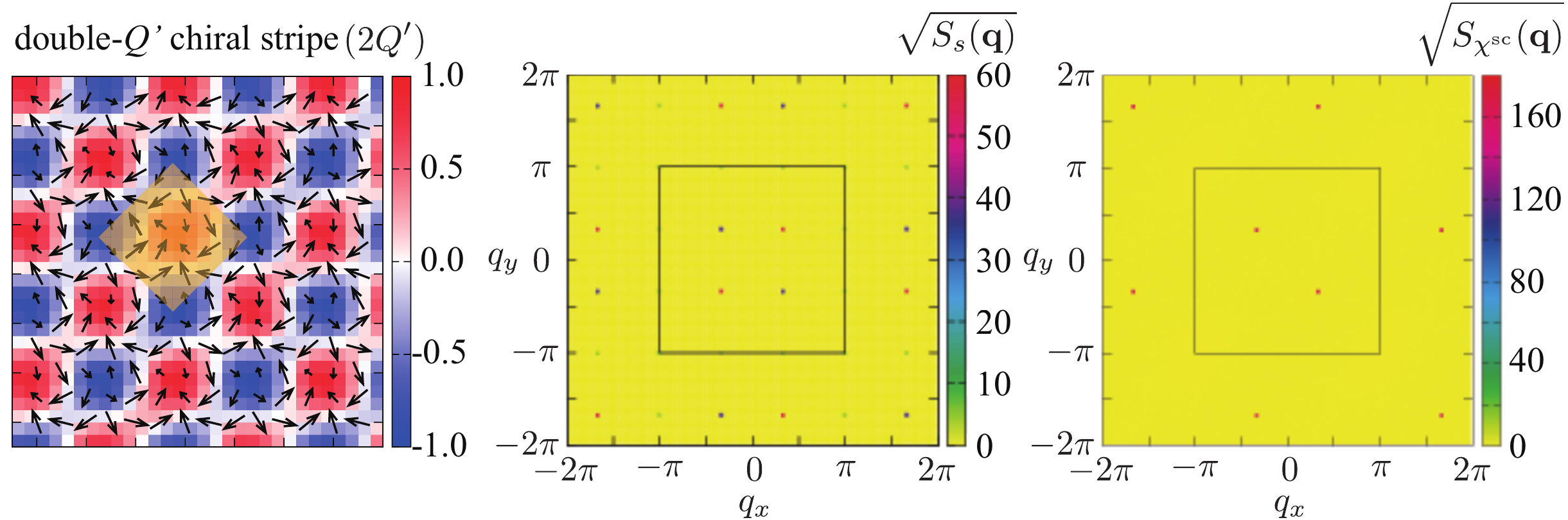} 
\caption{
\label{Fig:spinconfig_squ}
(Left) Snapshot of the spin configurations obtained by Monte Carlo simulations in the double-$Q'$ chiral stripe phase for $K=0.2$ on the square lattice. 
A part of the whole lattice with $N=96^2$ is shown.  
The contour shows the $z$ component of the spin moment, and the arrows represent the $xy$ components. 
The yellow square represents the magnetic unit cell. 
(Middle)
Square root of the corresponding spin structure factor.  
(Right) 
Square root of the corresponding chirality structure factor.  
In the middle and right figures, the squares represent the first Brillouin zone for the square lattice.
}
\end{center}
\end{figure*}

First, we show the result on the square lattice. 
In order to investigate the magnetic instability, we perform Monte Carlo simulations at sufficiently low temperature, $T=0.01$, with using the simulated annealing described in the previous section.  
We set the ordering vectors as $\mathbf{Q}_1=(2\pi/6, 2\pi/6)$ and $\mathbf{Q}_2=R(\pi/2)\mathbf{Q}_1$ in the model in Eq.~(\ref{eq:effHam_spin}) defined on the square lattice, although the following results are qualitatively unchanged for other sets of symmetry-related $\mathbf{Q}_\nu$. 

Figure~\ref{Fig:Kdep_phase_squ}(a) shows $K$ dependences of the $\mathbf{Q}_\nu$ component of the magnetization, $m_{\mathbf{Q}_\nu}$. 
At $K=0$, the single-$Q$ helical state is realized because Eq.~(\ref{eq:effHam_spin}) is reduced to the RKKY interaction in Eq.~(\ref{eq:RKKYHam}). 
Indeed, our Monte Carlo result indicates that $m_{\mathbf{Q}_1}$ becomes nonzero, while $m_{\mathbf{Q}_2}$ vanishes (the state with $m_{\mathbf{Q}_1}=0$ and $m_{\mathbf{Q}_2}\neq 0$ is also obtained depending on the initial conditions and the scheduling in the simulated annealing). 
Once we turn on $K$, the $\mathbf{Q}_2$ component becomes nonzero and develops as increasing $K$. This indicates that the single-$Q$ state has an instability toward the double-$Q$ state for an infinitesimal $K$. 
With increasing $K$, $m_{\mathbf{Q}_1}$ decreases and $m_{\mathbf{Q}_2}$ increases, and they gradually approaches the same value, while they take different values, at least, in the calculated range for $K<1.2$. 

We present a Monte Carlo snapshot of the spin configuration of this double-$Q$ state in the left panel of Fig.~\ref{Fig:spinconfig_squ}. The spin configuration is clearly different from the single-$Q$ helical state and modulated by the second $Q$ component. 
The spin structure factor in the middle panel of Fig.~\ref{Fig:spinconfig_squ} shows two pairs of Bragg peaks at $\pm\mathbf{Q}_1$ and $\pm\mathbf{Q}_2$ with different intensities. 

On the other hand, in terms of the spin scalar chirality, the double-$Q$ state is accompanied by the chiral density wave only with the wave number $\mathbf{Q}_2$, as shown in Fig.~\ref{Fig:Kdep_phase_squ}(b).  
Indeed, the structure factor shown in the right panel of Fig.~\ref{Fig:spinconfig_squ} has a peak only at $\pm \mathbf{Q}_2$. 

We summarize the phase diagram in Fig.~\ref{Fig:Kdep_phase_squ}(c). 
As mentioned above, the single-$Q$ helical state is realized only at $K=0$, and the double-$Q$ state with chirality stripe is stable for all $K>0$: the critical value of $K$ is $K_c=0$. 
We call this double-$Q$ state for $K>0$ ``double-$Q'$ chiral stripe", where $Q'$ means that the amplitudes of the $\mathbf{Q}_1$ and $\mathbf{Q}_2$ components are different. 

We find that this double-$Q'$ chiral stripe is similar to the double-$Q$ state recently found in the Kondo lattice model in Ref.~\onlinecite{Ozawa_doi:10.7566/JPSJ.85.103703}. The real-space spin structure was described in the form  
\begin{align}
\label{eq:doubleQ}
\mathbf{S}_i =
\left(
    \begin{array}{c}
     \sqrt{1-b^2+b^2 \cos \mathbf{Q}_2 \cdot \mathbf{r}_i}   \cos \mathbf{Q}_1 \cdot \mathbf{r}_i  \\
     b \sin \mathbf{Q}_2 \cdot \mathbf{r}_i \\
     \sqrt{1-b^2+b^2 \cos \mathbf{Q}_2 \cdot \mathbf{r}_i}  \sin \mathbf{Q}_1 \cdot \mathbf{r}_i  
          \end{array}
  \right)^{\rm T},
\end{align}
where $b$ represents the amplitude of the $\mathbf{Q}_2$ component; 
the superscript T denotes the transpose of the vector. 
As discussed in the previous study~\cite{Ozawa_doi:10.7566/JPSJ.85.103703}, the spin configuration in Eq.~(\ref{eq:doubleQ}) has two notable aspects. 
One is that it is continuously connected to the single-$Q$ helical state as $b\to 0$, which indicates that the double-$Q$ state is topologically trivial. 
The other is that the $\mathbf{Q}_2$ component is introduced in the perpendicular direction to the $\mathbf{Q}_1$ helical plane. 
Due to the latter, the energy loss from the higher harmonics at the RKKY level is small enough compared to the energy gain in the higher-order terms~\cite{Ozawa_doi:10.7566/JPSJ.85.103703}. 
We will discuss this point for our effective model in Sec.~\ref{sec:Energy comparison}. 

\subsubsection{Triangular lattice}
\label{sec:Triangular lattice}

\begin{figure}[htb!]
\begin{center}
\includegraphics[width=0.7 \hsize]{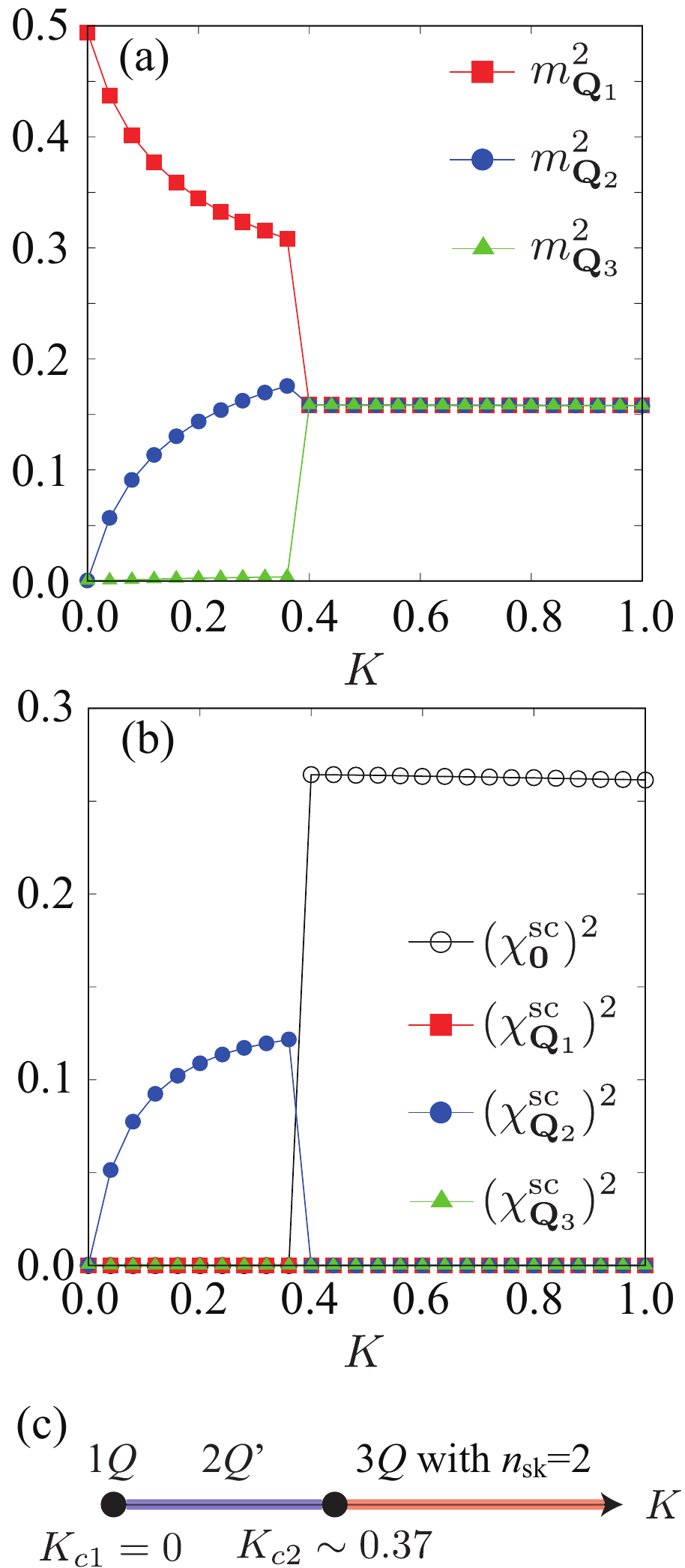} 
\caption{
\label{Fig:Kdep_phase_tri}
(a) $K$ dependences of the $\mathbf{Q}_\nu$ components of the squared magnetization obtained by Monte Carlo simulations for the model in Eq.~(\ref{eq:effHam_spin}).
The parameters are $\mathbf{Q}_1=(2\pi/6,0)$, $\mathbf{Q}_2=R(2\pi/3)\mathbf{Q}_1$ and $\mathbf{Q}_3=R(4\pi/3)\mathbf{Q}_1$. 
(b) $K$ dependences of the $\mathbf{Q}_\nu$ and uniform ($\mathbf{q}=0$) components of the squared scalar chirality for the same parameters in (a). 
(c) Phase diagram for the triangular lattice case. 
The single-$Q$ helical state ($1Q$) is stable only at $K=0$, the double-$Q'$ chiral stripe ($2Q'$) appears for $0<K\lesssim 0.37$, and the triple-$Q$ $n_{\rm sk}=2$ skyrmion crystal ($3Q$ with $n_{\rm sk}=2$) is realized for $K\gtrsim 0.37$. 
}
\end{center}
\end{figure}

\begin{figure*}[htb!]
\begin{center}
\includegraphics[width=1.0 \hsize]{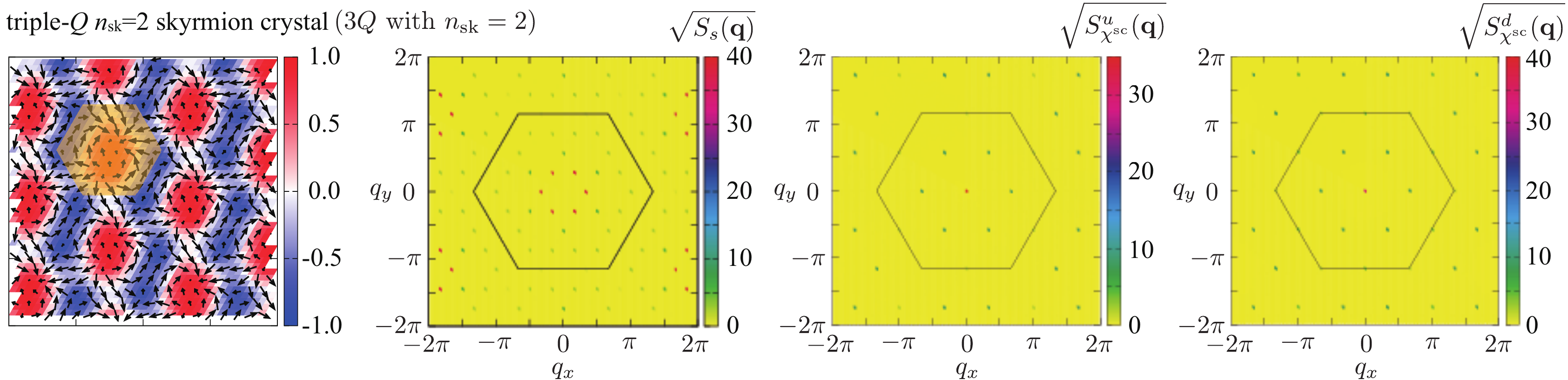} 
\caption{
\label{Fig:spinconfig_tri}
(Leftmost) Snapshot of the spin configurations obtained by Monte Carlo simulations in the triple-$Q$ $n_{\rm sk}=2$ skyrmion crystal phase for $K=0.48$ on the triangular lattice. 
A part of the whole lattice with $N=96^2$ is shown. 
The contour shows the $z$ component of the spin moment, and the arrows represent the $xy$ components. 
The yellow hexagon represents the magnetic unit cell. 
(Middle left)
Square root of the corresponding spin structure factor.  
(Middle right and rightmost) 
Square root of the corresponding chirality structure factors for the upward and downward triangles.  
In the right three figures, the hexagons represent the first Brillouin zone for the triangular lattice. 
}
\end{center}
\end{figure*}

Next, we study the phase diagram on the triangular lattice. 
We consider the specific set of ordering vectors $\mathbf{Q}_1=(2\pi/6,0)$, $\mathbf{Q}_2=R(2\pi/3)\mathbf{Q}_1$, and $\mathbf{Q}_3=R(4\pi/3)\mathbf{Q}_1$ in the model in Eq.~(\ref{eq:effHam_spin}) defined on the triangular lattice, while qualitative features are expected to hold for arbitrary sets of symmetry-related $\mathbf{Q}_\nu$, as in the square lattice case.
The obtained results of the magnetic moments and chirality are shown in Figs.~\ref{Fig:Kdep_phase_tri}(a) and \ref{Fig:Kdep_phase_tri}(b), respectively.  
In the spin sector, the single-$Q$ state at $K=0$ turns into a double-$Q$ state for $K>0$. 
We find that this double-$Q$ state corresponds to a straightforward extension of the double-$Q'$ chiral stripe obtained for the square lattice: the spin structure is expressed by the same form as Eq.~(\ref{eq:doubleQ}) with choosing two $\mathbf{Q}_\nu$ out of three [e.g., $\mathbf{Q}_1$ and $\mathbf{Q}_2$ in the solution obtained in Figs.~\ref{Fig:Kdep_phase_tri}(a) and \ref{Fig:Kdep_phase_tri}(b)]. 

While increasing $K$, however, we find another phase transition at $K \sim 0.37$. 
At the transition, the magnetic moment changes discontinuously, as shown in Fig.~\ref{Fig:Kdep_phase_tri}(a); 
all $m_{\mathbf{Q}_\nu}$ become nonzero for $K \gtrsim 0.37$, which indicates that the spin state is a triple-$Q$ order. 
There, the amplitudes of the $\mathbf{Q}_1$, $\mathbf{Q}_2$, and $\mathbf{Q}_3$ components are equivalent, as shown in Fig.~\ref{Fig:Kdep_phase_tri}(a). 
This is also clearly seen in the spin structure factor shown in Fig.~\ref{Fig:spinconfig_tri}.  
On the other hand, in the scalar chirality sector, the uniform ($\mathbf{q}=0$) component is induced, while the $\mathbf{Q}_\nu$ components all vanish, as shown in Figs.~\ref{Fig:Kdep_phase_tri}(b). 

Thus, the state for $K \gtrsim 0.37$ is the triple-$Q$ order with nonzero net scalar chirality, which is common to the skyrmion states studied  in noncentrosymmetric chiral magnets~\cite{rossler2006spontaneous,Yi_PhysRevB.80.054416,Binz_PhysRevLett.96.207202,nagaosa2013topological} and centrosymmetric frustrated magnets~\cite{Okubo_PhysRevLett.108.017206,leonov2015multiply,Lin_PhysRevB.93.064430,Hayami_PhysRevB.93.184413}. 
However, it is qualitatively distinct from the ordinary skyrmion crystals. 
Actually, we find that the triple-$Q$ state is the same as the unconventional skyrmion crystal discovered in Ref.~\onlinecite{ozawa2016zero}. 
The real-space spin structure is described by an equal superposition of three helices as 
\begin{align}
\label{eq:tripleQ}
\mathbf{S}_i =\frac{1}{N_i}
\left(
    \begin{array}{c}
    \cos \mathbf{Q}_1 \cdot \mathbf{r}_i  \\
     \cos \mathbf{Q}_2 \cdot \mathbf{r}_i  \\
    \cos \mathbf{Q}_3 \cdot \mathbf{r}_i
          \end{array}
  \right)^{\rm T},
\end{align}
where $N_i$ is the site-dependent renormalization factor to satisfy $|\mathbf{S}_i| =1$: $N_i = \sqrt{\sum_{\nu=1}^3 \cos^2 \mathbf{Q}_\nu \cdot \mathbf{r}_i}$. 
This triple-$Q$ state has mainly three different aspects from the conventional skyrmion crystals. 
The first point is that the topological number $n_{\rm sk}$ is two per magnetic unit cell, as shown in Fig.~\ref{Fig:spinconfig_tri}~\cite{ozawa2016zero}. 
This is in sharp contrast to the previous skyrmions with $n_{\rm sk}=1$. 
The second point is the symmetry in the ordered phase. 
The obtained triple-$Q$ state preserves inversion and spin rotational symmetries [O(3) symmetry], while most of the previous skyrmions in centrosymmetric systems have only U(1) symmetry because they are stabilized in an applied magnetic field. 
This peculiar symmetry may lead to new Goldstone modes.  
The third is the absence of the nonzero $\mathbf{Q}$ components in the scalar chirality, as shown in Fig.~\ref{Fig:Kdep_phase_tri}(b). This is in contrast to the presence of the six peaks in the chirality structure factor in the skyrmion crystal phase, e.g., found in Ref.~\onlinecite{Hayami_PhysRevB.93.184413}. 
Following Ref.~\onlinecite{ozawa2016zero}, we call this triple-$Q$ state the triple-$Q$ $n_{\rm sk}=2$ skyrmion crystal. 

We summarize the phase diagram in Fig.~\ref{Fig:Kdep_phase_tri}(c). 
As in the square lattice case, the phase boundary between the single-$Q$ helical state and the double-$Q'$ chiral stripe is at $K=K_{c1}=0$. 
In the triangular lattice case, however, there is another phase transition at $K=K_{c2} \sim 0.37$, above which the triple-$Q$ $n_{\rm sk}=2$ skyrmion crystal appears.

\subsection{Energy comparison}
\label{sec:Energy comparison}

We here discuss why the multiple-$Q$ orderings in Eqs.~(\ref{eq:doubleQ}) and (\ref{eq:tripleQ}) are realized in the presence of the biquadratic $K$ term, which mimics the fourth-order contribution in the Kondo lattice model. 
We discuss the energy comparison between the single-$Q$ helical state and the double-$Q'$ chiral stripe in Sec.~\ref{sec:Double-$Q'$ chiral stripe} and among the single-$Q$ helical state, the double-$Q'$ chiral stripe, and triple-$Q$ $n_{\rm sk}=2$ skyrmion crystal in Sec.~\ref{sec:Triple-$Q$ skyrmion crystal}. 
We also comment on a coplanar double-$Q$ state on the square lattice, which is the counterpart of the triple-$Q$ $n_{\rm sk}=2$ skyrmion crystal on the triangular lattice, in Sec.~\ref{sec:Double-$Q$ vortex crystal}. 

\subsubsection{Double-$Q'$ chiral stripe}
\label{sec:Double-$Q'$ chiral stripe}

First, we discuss the competition between the single-$Q$ helical and the double-$Q'$ chiral stripe states. 
We here show that the single-$Q$ helical state has higher energy than the double-$Q'$ chiral stripe for $0<K \ll \tilde{J}$ in our effective spin model in Eq.~(\ref{eq:effHam_spin}). 
The following argument is generic and applicable to both the square and triangular lattices. 
The argument is basically parallel to that for the original Kondo lattice model~\cite{Ozawa_doi:10.7566/JPSJ.85.103703}, which suggests that our effective spin model captures the instability toward the double-$Q'$ chiral stripe in the Kondo lattice model. 

In order to estimate the energy of the double-$Q'$ chiral stripe, we expand the square root in Eq.~(\ref{eq:doubleQ}) with respect to $b^2$: 
\begin{align}
&\sqrt{1-b^2+b^2 \cos^2 \bm{Q}_2 \cdot \bm{r}_i} \nonumber \\
&= 1+\frac{b^2}{4}(1-\cos 2 \bm{Q}_2 \cdot \bm{r}_i) 
 -\frac{b^4}{32} (1 - \cos 2 \bm{Q}_2 \cdot \bm{r}_i)^2  +\cdots \nonumber \\
& = C_0 +  C_2 \cos 2 \bm{Q}_2 \cdot \bm{r}_i + C_4 \cos 4 \bm{Q}_2 \cdot \bm{r}_i  + \cdots, 
\label{eq:sqrt_expansion}
\end{align}
where $C_{2l}$ ($l=0, 1, 2, \cdots$) are given by 
\begin{align}
C_{0} &= 1-\frac{b^2}{4}-\frac{3} {64}b^4  +\cdots,  \\
C_{2} &= \frac{b^2}{4} + \frac{b^4}{16} + \cdots, \\
C_4   &= -\frac{b^4}{64} - \cdots, 
\end{align}
and so on. 
Then, the spin pattern in Eq.~(\ref{eq:doubleQ}) is represented by 
\begin{align}
\mathbf{S}_i &= (C_0 + C_2 \cos 2 \mathbf{Q}_2 \cdot \mathbf{r}_i + C_4 \cos 4 \mathbf{Q}_2 \cdot \mathbf{r}_i) \nonumber \\ 
&\times ( \hat{\mathbf{x}}\cos \mathbf{Q}_1 \cdot \mathbf{r}_i+\hat{\mathbf{y}}\sin \mathbf{Q}_1 \cdot \mathbf{r}_i   ) \nonumber \\
&+ \hat{\mathbf{z}} b \sin \mathbf{Q}_2 \cdot \mathbf{r}_i +\mathcal{O}(b^6). 
\end{align}
By Fourier transformation, we obtain 
\begin{align}
\label{eq:SQ1}
\frac{1}{\sqrt{N}}\mathbf{S}_{\mathbf{Q_1}}&\simeq\hat{\mathbf{x}} \frac{C_0}{2}+i\hat{\mathbf{y}} \frac{C_0}{2}, \\
\label{eq:SQ2}
\frac{1}{\sqrt{N}}\mathbf{S}_{\mathbf{Q_2}}&\simeq\hat{\mathbf{z}} \frac{b}{2}, \\
\label{eq:SQ12Q2}
\frac{1}{\sqrt{N}}\mathbf{S}_{\mathbf{Q_1}+2\mathbf{Q}_2}&\simeq\hat{\mathbf{x}} \frac{C_2}{4}+i\hat{\mathbf{y}} \frac{C_2}{4}, \\
\label{eq:SQ12Q2_2}
\frac{1}{\sqrt{N}}\mathbf{S}_{\mathbf{Q_1}-2\mathbf{Q}_2}&\simeq\hat{\mathbf{x}} \frac{C_2}{4}+i\hat{\mathbf{y}} \frac{C_2}{4}.  
\end{align}
By substituting Eqs.~(\ref{eq:SQ1}) and (\ref{eq:SQ2}) into the effective spin model in Eq.~(\ref{eq:effHam_spin}), we obtain the energy per site for the double-$Q'$ chiral stripe: 
\begin{align}
E^{2Q'}&= -\tilde{J} \left( 1 -\frac{b^4}{32} \right) 
+ \frac{K}{2} \left(
 1-b^2 + \frac{7b^4}{16}
 \right) +\mathcal{O}(b^6), \\
 \label{eq:compare1Q2Q}
 &\sim E^{1Q}+\frac{\tilde{J}b^4}{32}-\frac{Kb^2}{2}. 
\end{align}
In the second line, we neglect the contributions $K b^4$ because we assume $\tilde{J} \gg  K$. 

The important observation in Eq.~(\ref{eq:compare1Q2Q}) is that there are no contributions proportional to $\tilde{J}b^2$. This comes from the fact that the modulation with the $\mathbf{Q}_2$ component is perpendicular to the helical plane with the $\mathbf{Q}_1$ component, as described in Sec.~\ref{sec:Magnetic phase diagram}. 
Obviously, the second term in Eq.~(\ref{eq:compare1Q2Q}) leads to the energy loss in the double-$Q'$ chiral stripe, which is consistent with the argument that the RKKY interaction favors the single-$Q$ helical state. 
On the other hand, the third term in Eq.~(\ref{eq:compare1Q2Q}) represents the energy gain in the double-$Q'$ chiral stripe. 
The condition to stabilize the double-$Q'$ chiral stripe, i.e., $E^{2Q'} < E^{1Q}$ reads 
\begin{align}
0< b^2 < \frac{16K}{\tilde{J}}. 
\end{align}
This means that an infinitesimal $K$ leads to the instability of the single-$Q$ helical state by introducing the second component with the amplitude $b$. 

We note that, strictly speaking, we need to take into account the contributions from Eqs.~(\ref{eq:SQ12Q2}) and (\ref{eq:SQ12Q2_2}) because they also give the energy in the order of $\tilde{J} b^4$. 
However, we can neglect such contributions when the bare susceptibility has considerably small amplitudes at $\mathbf{Q}_1 \pm 2 \mathbf{Q}_2$ compared to the distinct peaks at $\mathbf{Q}_1$ and $\mathbf{Q}_2$, as shown in Fig.~\ref{Fig:suscep}. 
The detailed description including the contributions of higher harmonics was given in Ref.~\onlinecite{Ozawa_doi:10.7566/JPSJ.85.103703}. 

\subsubsection{Triple-$Q$ skyrmion crystal}
\label{sec:Triple-$Q$ skyrmion crystal}

Next, we turn to the stability of the triple-$Q$ state on the triangular lattice, which appears for relatively large $K$. 
We can rewrite the effective Hamiltonian in Eq.~(\ref{eq:effHam_spin}) into the form 
\begin{align}
\label{eq:Ham_effspin_GL}
\mathcal{H}=&-2\tilde{J}(|S_{\mathbf{Q}_1}|^2+|S_{\mathbf{Q}_2}|^2+|S_{\mathbf{Q}_3}|^2) \nonumber \\
&+ 2\tilde{K} (|S_{\mathbf{Q}_1}|^2+|S_{\mathbf{Q}_2}|^2+|S_{\mathbf{Q}_3}|^2)^2 \nonumber \\
&- 4\tilde{K} (|S_{\mathbf{Q}_1}|^2 |S_{\mathbf{Q}_2}|^2+|S_{\mathbf{Q}_2}|^2 |S_{\mathbf{Q}_3}|^2+|S_{\mathbf{Q}_3}|^2 |S_{\mathbf{Q}_1}|^2), 
\end{align}
where $|S_{\mathbf{Q}_\nu}|^2=\mathbf{S}_{\mathbf{Q}_\nu} \cdot \mathbf{S}_{-\mathbf{Q}_\nu}$. 
From this form, for optimizing the energy with respect to $K=N\tilde{K}$, it is necessary to minimize the contribution from the second term in Eq.~(\ref{eq:Ham_effspin_GL}) and to maximize that from the third term. 
This is accomplished by the conditions: (i) the amplitude of the triple-$Q$ component is taken to be the same because of the sum rule $\sum_\nu |S_{\mathbf{Q}_\nu}|^2 = N$, and (ii) the directions of $\mathbf{Q}_\nu$ modulations are perpendicular to each other, namely, the magnetic moments for the $\mathbf{Q}_\nu$ components are perpendicular to one another. 
The latter condition indicates that the contribution of the higher harmonics is small, as discussed above for the double-$Q'$ state in the square lattice case. 
The spin configuration in Eq.~(\ref{eq:tripleQ}) satisfies these two conditions, and hence, it is chosen as the ground state in the large $K$ region. 

Let us evaluate the energy of the triple-$Q$ $n_{\rm sk}=2$ skyrmion crystal. 
Although the comparison with the double-$Q'$ chiral stripe is difficult because Eq.~(\ref{eq:compare1Q2Q}) is justified only for small $b$, we here compare the energy of the triple-$Q$ state with that of the single-$Q$ state. 
In the triple-$Q$ state in Eq.~(\ref{eq:tripleQ}), the magnetic moments for the $\mathbf{Q}_\nu$ components are given by 
\begin{align}
\label{eq:SQ_tripleQ}
|\mathbf{S}_{\mathbf{Q}_1} |= |\mathbf{S}_{\mathbf{Q}_2} |=| \mathbf{S}_{\mathbf{Q}_3} |= \sqrt{\frac{N}{6}}(1-\eta), 
\end{align}
where $\eta$ represents a small correction from the higher harmonics, e.g., $\mathbf{Q}_1+2\mathbf{Q}_2$ and $3\mathbf{Q}_1$, leading to the energy loss at the level of the RKKY interaction. 
By substituting Eq.~(\ref{eq:SQ_tripleQ}) into Eq.~(\ref{eq:effHam_spin}), we obtain the energy per site for the triple-$Q$ $n_{\rm sk}=2$ skyrmion crystal as 
\begin{align}
\label{eq:tripleQenergy}
E^{3Q}=-\tilde{J} (1-2 \eta) + \frac{K}{6} (1-4 \eta)+\mathcal{O}(\eta^2). 
\end{align}
Monte Carlo simulations give $\eta \simeq 0.05$ almost independent of $K$. 
From Eq.~(\ref{eq:tripleQenergy}), we find that the energy of the triple-$Q$ $n_{\rm sk}=2$ skyrmion crystal is higher than the single-$Q$ state for small $K$. 
While increasing $K$, however, the energy of the triple-$Q$ $n_{\rm sk}=2$ skyrmion crystal becomes lower for $K>3/11\tilde{J} \sim 0.27\tilde{J}$ when we set $\eta=0.05$. This is qualitatively consistent with the result by numerical simulations in Fig.~\ref{Fig:Kdep_phase_tri}(a), giving the phase boundary at $K_{c2} \sim 0.37 \tilde{J} $. 
The underestimate of $K_{c2}$ is reasonable as we neglect the double-$Q'$ chiral stripe in the current analysis. 

\subsubsection{Coplanar double-$Q$ state on the square lattice}
\label{sec:Double-$Q$ vortex crystal}

Let us comment on the reason why the second transition at $K_{c2}$ is absent in the square lattice case. 
We can think of a counterpart of the triple-$Q$ state in Eq.~(\ref{eq:tripleQ}) for the square lattice case, which may optimize the $K$ term along with the conditions discussed in Sec.~\ref{sec:Triple-$Q$ skyrmion crystal}.  As there are two $\mathbf{Q}_\nu$ in the square lattice case, such an optimal state will be given by 
\begin{align}
\label{eq:spindoulbeQflux}
\mathbf{S}_i =\frac{1}{N_i}
\left(
    \begin{array}{c}
    \cos \mathbf{Q}_1 \cdot \mathbf{r}_i  \\
     \cos \mathbf{Q}_2 \cdot \mathbf{r}_i  \\
    0
          \end{array}
  \right)^{\rm T},
\end{align}
where $N_i = \sqrt{\sum_{\nu=1}^2 \cos^2 \mathbf{Q}_\nu \cdot \mathbf{r}_i}$. 
Although the coplanar double-$Q$ state is expected to be stabilized for $K \gg \tilde{J}$ on the square lattice as it satisfies the two conditions in Sec.~\ref{sec:Triple-$Q$ skyrmion crystal}, it is not obtained as the lowest-energy state in our Monte Carlo simulations.    

Nonetheless, the coplanar double-$Q$ state sometimes appears as a metastable state when we perform simulated annealing with a sudden quench from high-temperature limit. 
We find that, although the energy of this coplanar double-$Q$ state is always higher than that in the double-$Q'$ chiral stripe, the difference becomes smaller for larger $K$; 
they are almost degenerate for $K \gg \tilde{J}$ within the numerical accuracy. 
This is because the energy loss in the $K$ term for the coplanar double-$Q$ state is almost similar to that for the double-$Q'$ chiral stripe when taking $b\sim 1$. 
The situation is in contrast to the triangular lattice case. 
The energy loss in the $K$ term for the triple-$Q$ $n_{\rm sk}=2$ skyrmion crystal is given by $(K/6) (1-4 \eta)$ in Eq.~(\ref{eq:tripleQenergy}), which is smaller than $(K/4) (1-4\eta)$ for the double-$Q'$ chiral stripe and coplanar double-$Q$ states. 
Moreover, from the numerical simulations, we find that the effect of higher harmonics in the coplanar double-$Q$ state, mainly from $\pm\mathbf{Q}_1 \pm2\mathbf{Q}_2$ and $\pm 3 \mathbf{Q}_1$, becomes larger compared to that in the double-$Q'$ chiral stripe. 
Thus, the coplanar double-$Q$ state is not stabilized in the effective model in Eq.~(\ref{eq:effHam_spin}) on the square lattice, and hence, there is no second transition for this case.

\section{Comparison to Itinerant model}
\label{sec:Comparison to Itinerant model}
\begin{figure}[htb!]
\begin{center}
\includegraphics[width=0.9 \hsize]{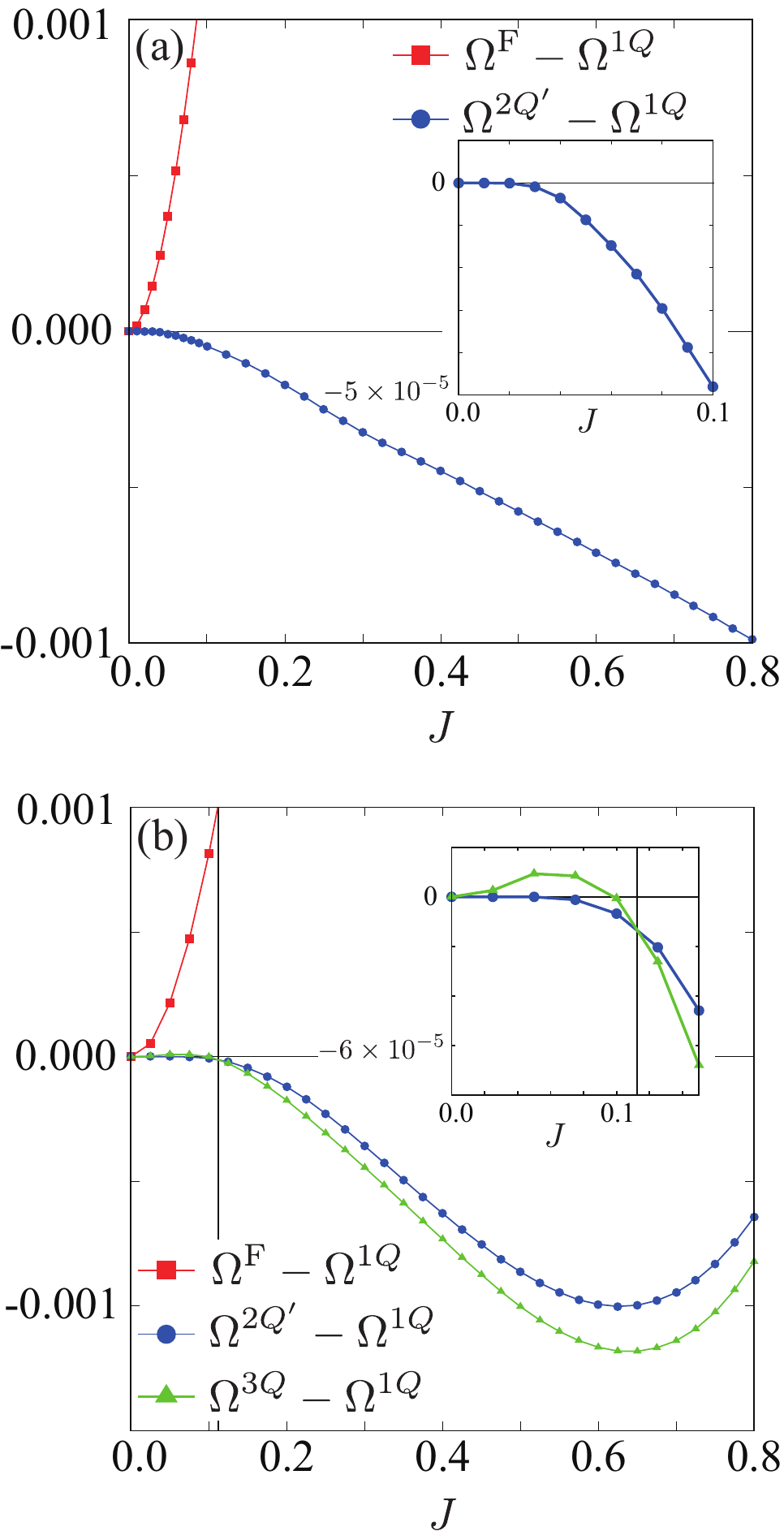} 
\caption{
\label{Fig:variational_energy}
(a) $J$ dependences of the grand potential for the ferromagnetic state (F) and the double-$Q'$ chiral stripe ($2Q'$) measured from that for the single-$Q$ helical state ($1Q$) on the square lattice at $t_3=-0.5$ and $\mu=0.98$. 
(b) $J$ dependences of the grand potential for the ferromagnetic state (F), the double-$Q'$ chiral stripe ($2Q'$), and the triple-$Q$ $n_{\rm sk}=2$ skyrmion crystal ($3Q$) measured from that for the single-$Q$ helical state ($1Q$) on the triangular lattice at $t_3=-0.85$ and $\mu=-3.5$. 
The inset of each figure shows the enlarged view for the small $J$ region. 
The vertical lines in (b) represent the phase boundary between the $2Q'$ and $3Q$ states. 
}
\end{center}
\end{figure}

So far, we have investigated the phase diagram of the spin model in Eq.~(\ref{eq:effHam_spin}), which we propose as the effective model for the Kondo lattice model in Eq.~(\ref{eq:Ham}) in the small $J$ region. 
In this section, we examine the validity of the effective model by comparing the obtained phase diagrams with those by the direct calculations for the Kondo lattice model. 

\subsection{Variational calculations}
\label{sec:Variational calculations}

In order to confirm the validity of the effective spin model obtained from the perturbative argument, we examine the ground state of the original Kondo lattice model in Eq.~(\ref{eq:Ham}). 
We here perform a variational calculation as follows: 
we compare the grand potential at zero temperature $\Omega= E-\mu n_{\rm e}$ for variational states with different magnetic orders in the localized spins, where $E=\langle \mathcal{H} \rangle /N$ is the internal energy per site, and $n_{\rm e}=\sum_{i\sigma} \langle c_{i\sigma}^\dagger c_{i\sigma} \rangle /N$ is the electron density. 
For the variational states, we assume the spin configurations given in Eqs.~(\ref{eq:spin_helical}), (\ref{eq:doubleQ}), and (\ref{eq:tripleQ}) with the same wave numbers used in the effective spin model. 
In the double-$Q'$ state in Eq.~(\ref{eq:doubleQ}), we deal with $b$ as the variational parameter. 
For comparison, we also calculate the grand potential for the ferromagnetic state, which is expected for large $J$ by the double-exchange mechanism~\cite{Zener_PhysRev.82.403,anderson1955considerations}; the spin configuration is given by $\mathbf{S}_i=(0,0,1)$. 
We consider these variational states in the system with $N=480^2$ under the periodic boundary conditions. 

Figure~\ref{Fig:variational_energy}(a) shows $J$ dependences of the grand potential at $t_3=-0.5$ and $\mu=0.98$ on the square lattice for the ferromagnetic state and the double-$Q'$ chiral stripe measured from that for the single-$Q$ helical state. The similar calculations were done in Ref.~\onlinecite{Ozawa_doi:10.7566/JPSJ.85.103703}. 
The results are consistent with those obtained in the effective spin model in Sec.~\ref{sec:Multiple-$Q$ magnetic instability}. 
The grand potential for the double-$Q'$ chiral stripe state is lower than that for the single-$Q$ helical state as well as the ferromagnetic state for all $J>0$ [see also the inset of Fig.~\ref{Fig:variational_energy}(a) for small $J$]. 

The results for the triangular lattice are also consistent with those for the effective spin model. 
Figure~\ref{Fig:variational_energy}(b) shows $J$ dependences of the grand potential at $t_3=-0.85$ and $\mu=-3.5$ measured from that for the single-$Q$ helical state. 
The double-$Q'$ chiral stripe 
becomes the lowest-energy state for $0<J\lesssim 0.11$, while the triple-$Q$ $n_{\rm sk}=2$ skyrmion crystal state replaces it for $J \gtrsim 0.11$. 

Thus, for both square and triangular lattice cases, the variational calculations indicate that the multiple-$Q$ states are stable for nonzero $J$ in the Kondo lattice model. 
The sequence of the phases is consistent with that in the effective spin model in Eq.~(\ref{eq:effHam_spin}) shown in Figs.~\ref{Fig:Kdep_phase_squ}(c) and \ref{Fig:Kdep_phase_tri}(c). 
The results strongly support that the effective model captures the instabilities toward the multiple-$Q$ states inherent to the Kondo lattice model.

\subsection{Higher-order corrections}
\label{sec:Higher-order corrections}

\begin{figure}[htb!]
\begin{center}
\includegraphics[width=0.9\hsize]{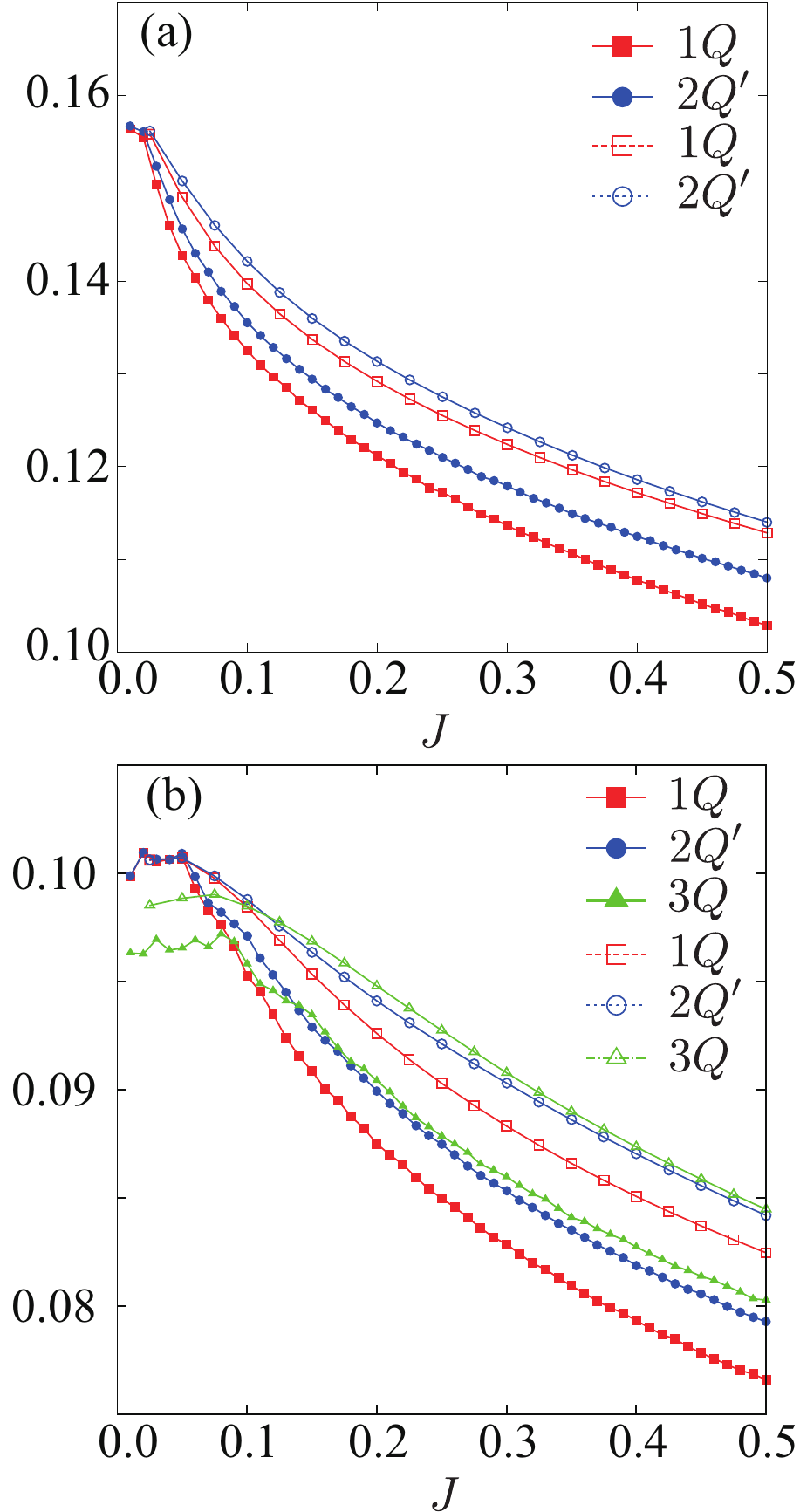} 
\caption{
\label{Fig:rpa}
Comparison between the free energy obtained from Eq.~(\ref{eq:renofreeenergy}) at $T=10^{-5}$ (closed symbols) and the grand potential at $T=0$ by the variational calculations (open symbols): 
(a) the single-$Q$ helical state (1$Q$) and the double-$Q'$ chiral stripe (2$Q'$) on the square lattice ($t_3=-0.5$ and $\mu=0.98$) and (b) the single-$Q$ helical state, the double-$Q'$ chiral stripe, and the triple-$Q$ $n_{\rm sk}=2$ skyrmion crystal (3$Q$) on the triangular lattice ($t_3=-0.85$ and $\mu=-2.5$). 
}
\end{center}
\end{figure}

Our effective spin model in Eq.~(\ref{eq:effHam_spin}) includes the two contributions inferred from the perturbative expansion of the Kondo lattice model in terms of $J$. One is the bilinear $\tilde{J}$ term representing the expansion terms favoring a single-$Q$ helical order, whose dominant contribution is from the second-order RKKY interaction in Eq.~(\ref{eq:RKKYHam_G}). The other is the biquadratic $\tilde{K}$ term representing the expansion terms favoring a multiple-$Q$ noncoplanar order, whose most dominant contribution is in the fourth-order terms, Eq.~(\ref{eq:F42}). 
On the other hand, as discussed in Sec.~\ref{sec:Higher-order contributions}, each term in the dominant contributions in the expansion, $(\mathbf{S}_{\mathbf{Q}_\nu} \cdot \mathbf{S}_{-\mathbf{Q}_\nu})^n$ [Eq.~(\ref{eq:F2nessential}) and Fig.~\ref{Fig:diagram_rpa}] diverges in the low-temperature limit, even though the sum up to infinite order in Eq.~(\ref{eq:renofreeenergy}) converges. 
To reinforce the validity of our effective model formally including only the terms with $n=1$ and $2$, 
we here show that the infinite sum in Eq.~(\ref{eq:renofreeenergy}) does not change the qualitative phase diagram on the square and triangular lattices obtained in Sec.~\ref{sec:Multiple-$Q$ magnetic instability}. 

Figure~\ref{Fig:rpa} shows $J$ dependences of $-F_{(\mathbf{Q},-\mathbf{Q})} /(2J^2)$ in Eq.~(\ref{eq:renofreeenergy}) for the single-$Q$ helical and the double-$Q'$ chiral stripe on the square lattice [Fig.~\ref{Fig:rpa}(a)] and for the single-$Q$ helical, the double-$Q'$ chiral stripe, and the triple-$Q$ $n_{\rm sk}=2$ skyrmion crystal on the triangular lattice [Fig.~\ref{Fig:rpa}(b)]. 
The parameters are taken to be the same as Fig.~\ref{Fig:variational_energy}. 
For comparison, we also plot the $-(\Omega-\Omega^{(0)})/(2J^2)$ obtained by the variational calculations, where $\Omega^{(0)}$ represents the grand potential for $J=0$. 
As shown in Fig.~\ref{Fig:rpa}, the free energy including the higher-order dominant contributions also reproduces well the phase diagrams on both square and triangular lattices. 
The agreement supports the validity of our effective model including only the bilinear and biquadratic interactions in momentum space.

\section{Effect of magnetic field}
\label{sec:Effect of magnetic field}

We have found that the effective spin model in Eq.~(\ref{eq:effHam_spin}) captures the magnetic instability in the Kondo lattice model in Eq.~(\ref{eq:Ham}). 
The effective model is useful for exploration of further exotic magnetic orderings, as its simplicity allows us to bypass laborious calculations necessary for itinerant electron models. 
In this section, we extend our study to construct the magnetic phase diagram by incorporating the effect of an external magnetic field on the effective spin model. 
For this purpose, adding the Zeeman coupling term to the Hamiltonian in Eq.~(\ref{eq:effHam_spin}), we consider the Hamiltonian given by 
\begin{align}
\label{eq:zeeman}
\mathcal{H}=2\sum_\nu
\left[ -\tilde{J}\mathbf{S}_{\mathbf{Q_{\nu}}} \cdot \mathbf{S}_{-\mathbf{Q_{\nu}}}
+\tilde{K} (\mathbf{S}_{\mathbf{Q_{\nu}}} \cdot \mathbf{S}_{-\mathbf{Q_{\nu}}})^2 \right] -H \sum_i S_i^z. 
\end{align}
We discuss the magnetic phase diagrams under the magnetic field on the square lattice in Sec.~\ref{sec:Square lattice Mag} and the triangular lattice in Sec.~\ref{sec:Triangular lattice Mag}. 

\subsection{Square lattice}
\label{sec:Square lattice Mag}

\begin{figure}[htb!]
\begin{center}
\includegraphics[width=1.0 \hsize]{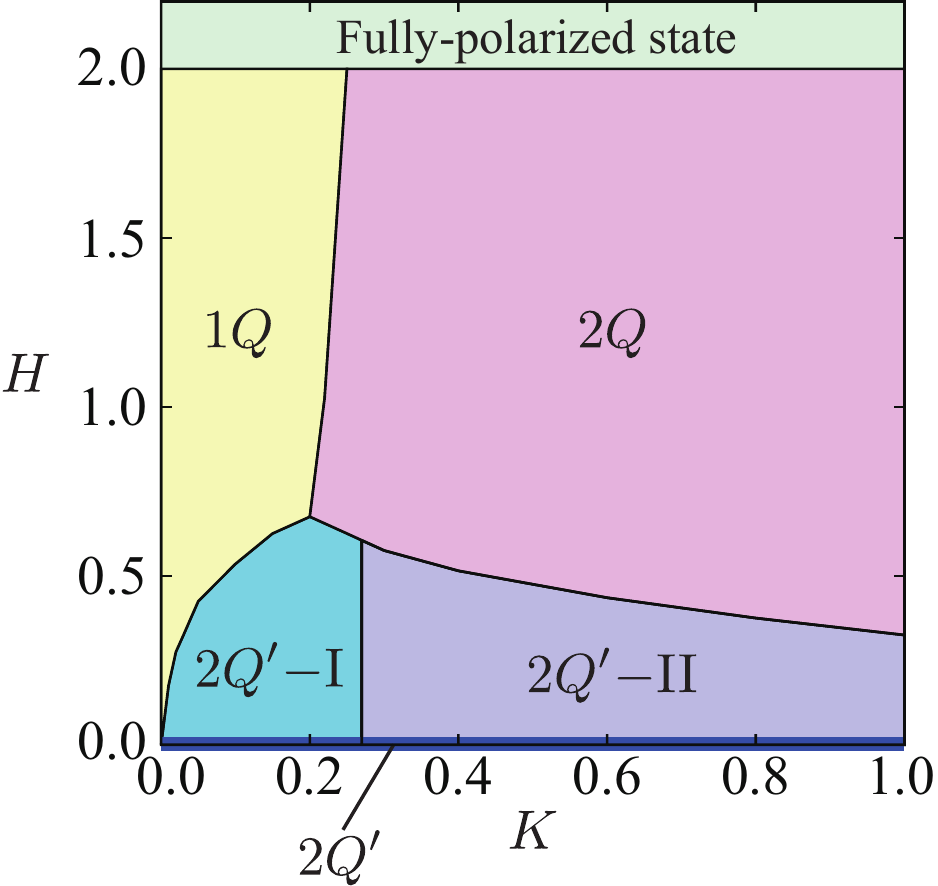} 
\caption{
\label{Fig:squ_phasediagram}
Phase diagram for the model in Eq.~(\ref{eq:zeeman}) on the square lattice with $\mathbf{Q}_1=(2\pi/6, 2\pi/6)$ and $\mathbf{Q}_2=R(\pi/2)\mathbf{Q}_1$. 
The spin configuration in each phase is shown in Fig.~\ref{Fig:squ_eachphase}. 
The phase diagram at $H=0$ coincides with that in Fig.~\ref{Fig:Kdep_phase_squ}(c). 
}
\end{center}
\end{figure}

\begin{figure*}[htb!]
\begin{center}
\includegraphics[width=1.0 \hsize]{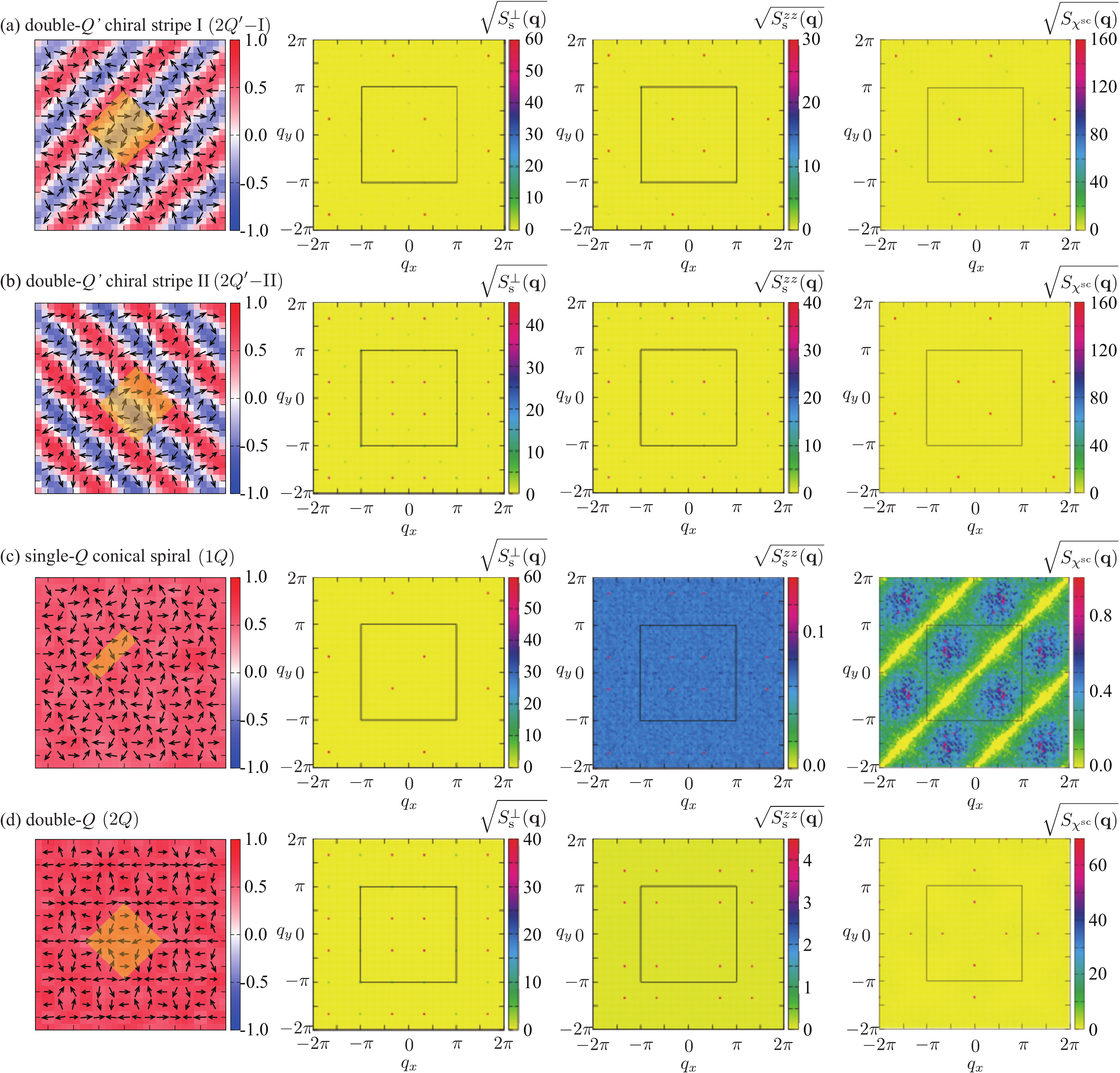} 
\caption{
\label{Fig:squ_eachphase}
(Leftmost) Snapshots of the spin configurations in (a) double-$Q'$ chiral stripe I ($2Q'$-I) for $K=0.1$ and $H=0.2$, (b) double-$Q'$ chiral stripe II ($2Q'$-II) for $K=0.4$ and $H=0.2$, (c) single-$Q$ conical state ($1Q$) for $K=0.1$ and $H=1.0$, and (d) double-$Q$ state ($2Q$) for $K=0.4$ and $H=1.0$, for  $\mathbf{Q}_1=(2\pi/6, 2\pi/6)$ and $\mathbf{Q}_2=R(\pi/2)\mathbf{Q}_1$. 
The contour shows the $z$ component of the spin moment. 
The yellow squares in (a), (b), (d) and rectangle in (c) represent the magnetic unit cells in each case. 
(Middle left and right) The square root of the $xy$ and $z$ components of the spin structure factor, respectively. 
Note that the $\mathbf{q}=0$ component is removed from $S^{zz}_s (\mathbf{q})$ to clearly show the structures other than $\mathbf{q}=0$. 
(Rightmost) The square root of the chirality structure factors. 
In the right three columns, the squares with a solid line show the first Brillouin zone. 
}
\end{center}
\end{figure*}

\begin{figure}[htb!]
\begin{center}
\includegraphics[width=0.9\hsize]{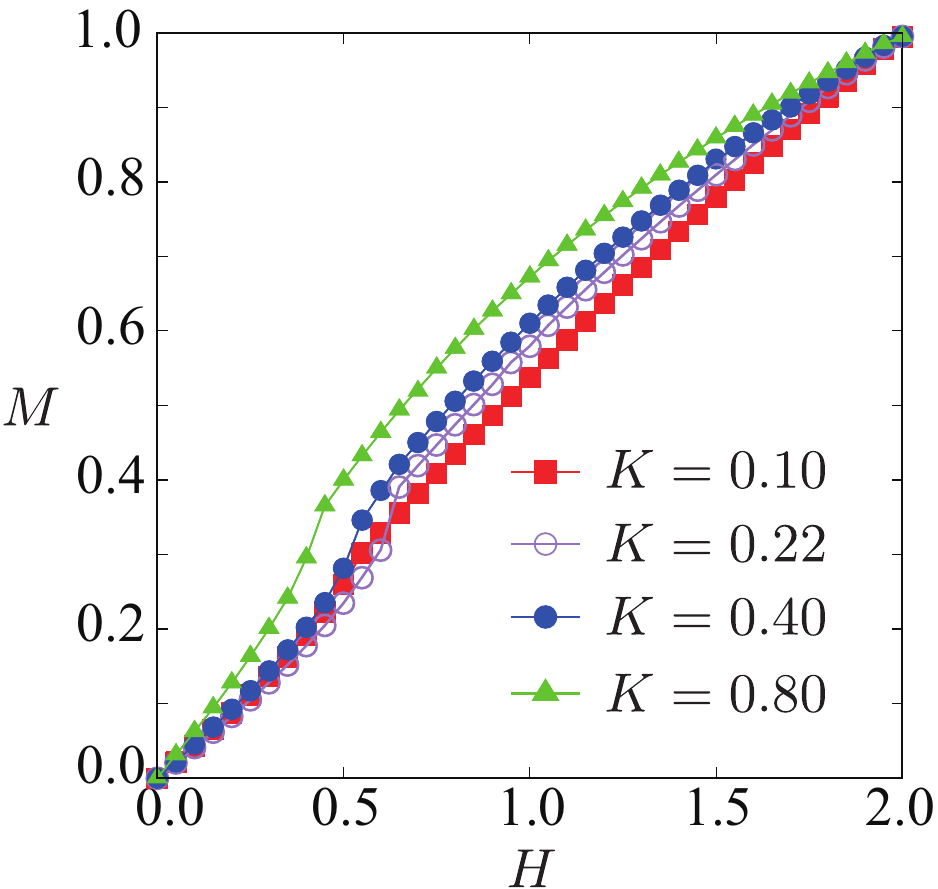} 
\caption{
\label{Fig:squ_mag}
Magnetization curves for $K=0.10$, $0.22$, $0.40$, and $0.80$. 
}
\end{center}
\end{figure}

Figure~\ref{Fig:squ_phasediagram} shows the $K$-$H$ phase diagram for the square lattice case obtained by our Monte Carlo simulations. We find four different phases other than the fully-polarized state in a large magnetic field. 
The phase boundaries are determined by analyzing the structure factors from Eqs.~(\ref{eq:spinstructurefactor}) to (\ref{eq:chiralstructurefactor}). 
All the phases show no net scalar chirality. 
In what follows, we describe the details of each phase one by one. We also discuss the magnetization processes. 

\paragraph{double-$Q'$ chiral stripe I (2$Q'$-I)}
This phase occupies a region at low $H$ and $K$. 
The spin configuration is characterized by the double-$Q$ noncoplanar modulation with different intensities, similar to the double-$Q'$ chiral stripe at $H=0$ discussed in Fig.~\ref{Fig:spinconfig_squ}, but the magnetic field breaks the O(3) symmetry in the present state; see the typical spin configuration in Fig.~\ref{Fig:squ_eachphase}(a). 
The $xy$ component of the spin structure factor has a single-$Q$ peak at $\mathbf{Q}_1$, while the $z$ component exhibits two peaks at $\mathbf{Q}_1$ and $\mathbf{Q}_2$ with different intensities, as shown in Fig.~\ref{Fig:squ_eachphase}(a). 
The chirality structure factor shows a single-$Q$ peak at $\mathbf{Q}_2$, similar to the double-$Q'$ chiral stripe at $H=0$. 

\paragraph{double-$Q'$ chiral stripe II (2$Q'$-II)}
This state occupies the low-$H$ and high-$K$ region of the phase diagram in Fig.~\ref{Fig:squ_phasediagram} and appears next to the double-$Q'$ chiral stripe I phase with increasing $K$. 
The spin configuration also resembles a double-$Q'$ chiral stripe state at $H=0$ in Fig.~\ref{Fig:spinconfig_squ}, whereas not only $z$ but also $xy$ components of the spin structure factor are characterized by the double-$Q$ peak structure in the present case, as shown in Fig.~\ref{Fig:squ_eachphase}(b). 
On the other hand, the chirality structure factor shows a single-$Q$ peak at $\mathbf{Q}_2$, similar to the chiral stripe at $H=0$. 
The transition from the double-$Q'$ I to II is presumably caused by the tendency that larger $K$ favors a larger amplitude of the double-$Q$ components, as already seen in the absence of $H$, as shown in Fig.~\ref{Fig:Kdep_phase_squ}(a). 

\paragraph{single-$Q$ conical state (1$Q$)}
This state occupies a region at higher $H$ and lower $K$, appearing next to the double-$Q'$ chiral stripe I phase with increasing $H$. 
This is a single-$Q$ conical state, whose spin configuration is given by $\mathbf{S}_i=(\sin \theta \cos \mathbf{Q}_\nu \cdot \mathbf{r}_i, \sin \theta \sin \mathbf{Q}_\nu \cdot \mathbf{r}_i, \cos \theta)$ with $\theta=\cos^{-1}(H/H_{\rm sat})$ where $H_{\rm sat}=2\tilde{J}$, independent of the value of $K$ [see Fig.~\ref{Fig:squ_eachphase}(c)]. 
We note that the conical state is realized even at $K=0$, where the model is reduced to a frustrated Heisenberg model in a magnetic field. 
With increasing $H$, the conical angle $\theta$ becomes smaller, and this phase continuously changes into the fully-polarized state at $H=2$, as shown in Fig.~\ref{Fig:squ_phasediagram}. 

\paragraph{double-$Q$ state (2$Q$)}
This phase occupies the largest portion of the phase diagram in Fig.~\ref{Fig:squ_phasediagram}, which is found next to the single-$Q$ conical state upon increasing $K$. 
The spin configuration is similar to the double-$Q$ state in Eq.~(\ref{eq:spindoulbeQflux}), although the present state shows a noncoplanar spin structure by spin canting along the magnetic field direction. It consists of a periodic array of vortices and antivortices, as shown in Fig.~\ref{Fig:squ_eachphase}(d). 
The $xy$ component of the spin structure factor shows the double-$Q$ peaks with equal intensities, while the $z$ components are negligibly small, except for a large peak at $\mathbf{q}=0$ from the net moment induced by the magnetic field [omitted in Fig.~\ref{Fig:squ_eachphase}(d) for clarity].  
With increasing $H$, the phase continuously changes into the fully-polarized state at $H=2$, as the single-$Q$ conical state. 
Note that this double-$Q$ state is also similar to the one reported on the triangular lattice in classical~\cite{Hayami_PhysRevB.94.174420} and quantum~\cite{Kamiya_PhysRevX.4.011023,Marmorini2014} magnets.

\paragraph{Magnetization curves}
Figure~\ref{Fig:squ_mag} shows the magnetization curves at several $K$. 
The magnetization is given by $M = m_{\mathbf{q}=0} = \sqrt{S_s(\mathbf{q}=0)/N}$ [see Eq.~(\ref{eq:m_q})]. 
For infinitesimal $H$, the magnetization becomes nonzero and continuously increases with increasing $H$ irrespective of $K$. 
This indicates continuous phase transitions at $H=0$. 
With further increasing $H$, the magnetization shows a continuous change between the double-$Q'$ chiral stripe I and the single-$Q$ conical state for small $K$ and between the double-$Q'$ chiral stripes I, II and the double-$Q$ state for large $K$. 
The result indicates that the transition is of second-order, although a more careful finite-size scaling is required to settle this point, especially, for large $K$. 
While further increasing $H$, the magnetization in both single-$Q$ conical and double-$Q$ states smoothly connect to the saturation value $M=1$ in the fully-polarized state at $H=2$.

\subsection{Triangular lattice}
\label{sec:Triangular lattice Mag}

\begin{figure}[htb!]
\begin{center}
\includegraphics[width=1.0 \hsize]{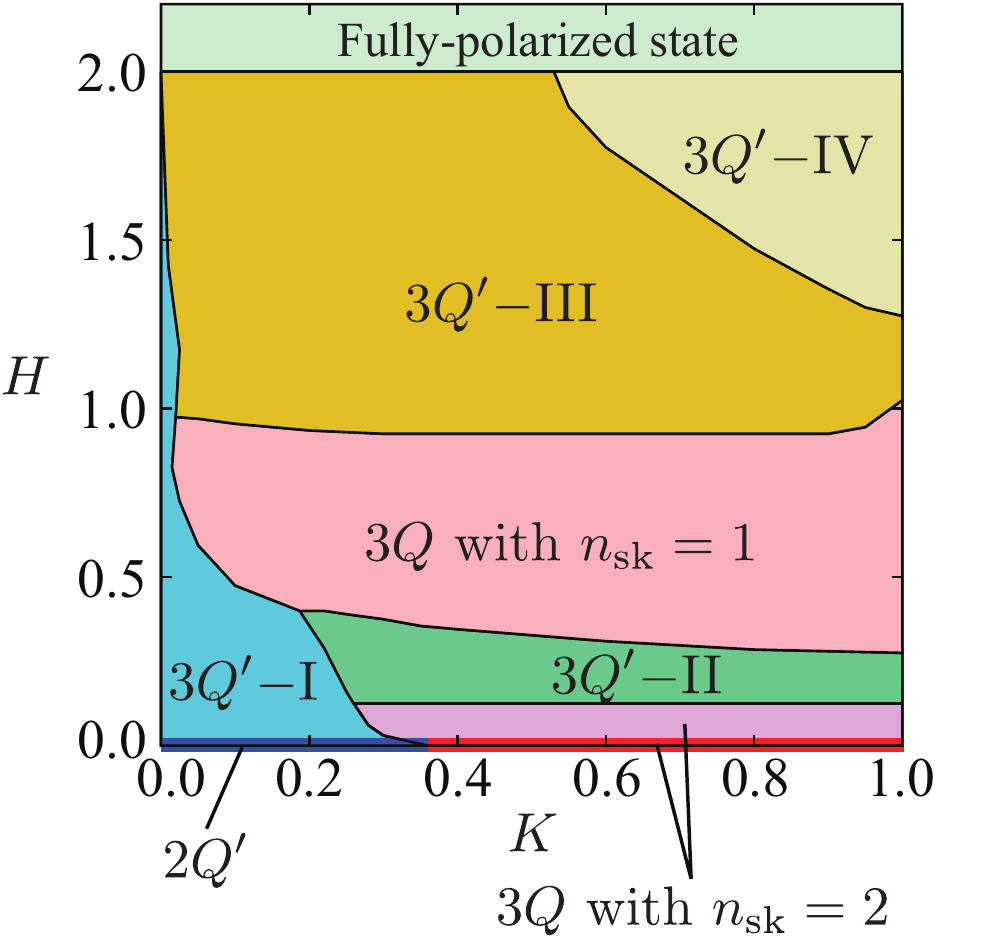} 
\caption{
\label{Fig:tri_phasediagram}
Phase diagram for the model in Eq.~(\ref{eq:zeeman}) on the triangular lattice with $\mathbf{Q}_1=(2\pi/6, 0)$, $\mathbf{Q}_2=R(2\pi/3)\mathbf{Q}_1$, and $\mathbf{Q}_3=R(4\pi/3)\mathbf{Q}_1$. 
The spin configuration in each phase is shown in Fig.~\ref{Fig:tri_eachphase}. 
The phase diagram at $H=0$ coincides with that in Fig.~\ref{Fig:Kdep_phase_tri}(c). 
}
\end{center}
\end{figure}

\begin{figure*}[htb!]
\begin{center}
\includegraphics[width=0.9 \hsize]{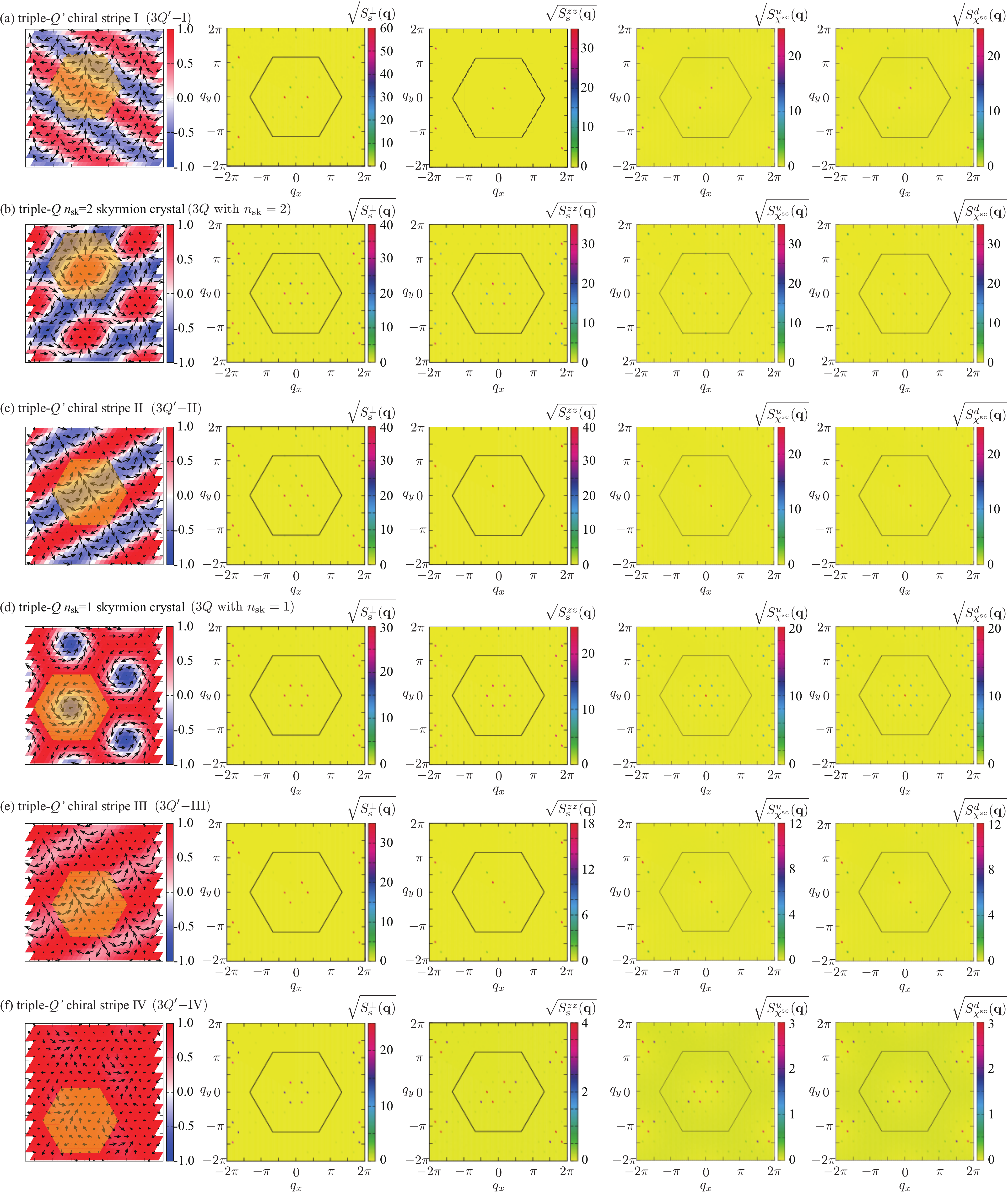} 
\caption{
\label{Fig:tri_eachphase}
(Leftmost) 
Snapshots of the spin configurations in (a) triple-$Q'$ chiral stripe I ($3Q'$-I) for $K=0.1$ and $H=0.2$, (b) triple-$Q$ $n_{\rm sk}=2$ skyrmion crystal for $K=0.4$ and $H=0.2$, (c) triple-$Q'$ chiral stripe ($3Q'$-II) for $K=0.6$ and $H=0.2$, (d) triple-$Q$ $n_{\rm sk}=1$ skyrmion crystal for $K=0.6$ and $H=0.8$, (e) triple-$Q'$ chiral stripe ($3Q'$-III) for $K=0.6$ and $H=1.2$, and (f) triple-$Q'$ chiral stripe ($3Q'$-IV) for $K=1.0$ and $H=1.6$, for  $\mathbf{Q}_1=(2\pi/6, 0)$, $\mathbf{Q}_2=R(2\pi/3)\mathbf{Q}_1$, and $\mathbf{Q}_3=R(4\pi/3)\mathbf{Q}_1$. 
The contour shows the $z$ component of the spin moment. 
The yellow hexagons show the magnetic unit cells. 
(Middle left and middle) The square root of the $xy$ and $z$ components of the spin structure factor, respectively. 
Note that the $\mathbf{q}=0$ component is removed from $S^{zz}_s (\mathbf{q})$ to clearly show the peaks at $\mathbf{q}\neq 0$. 
(Middle right and rightmost) The square root of the chirality structure factors for up and down triangles, respectively. 
In the right four columns, the hexagons with a solid line show the first Brillouin zone. 
}
\end{center}
\end{figure*}

\begin{figure}[htb!]
\begin{center}
\includegraphics[width=0.9 \hsize]{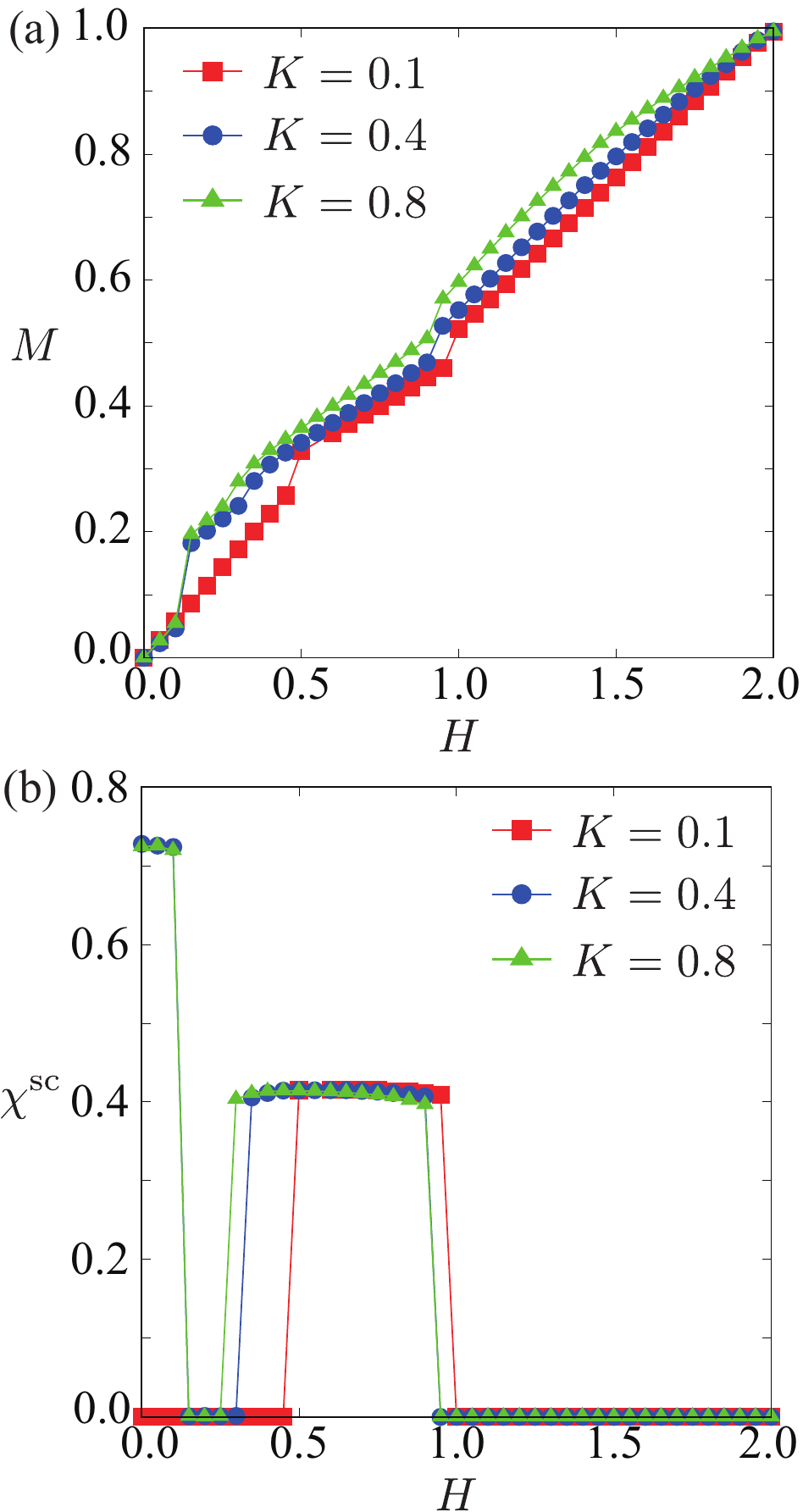} 
\caption{
\label{Fig:tri_mag}
(a) Magnetization curves and (b) the net scalar chirality for $K=0.1$, $0.4$, and $0.8$. 
}
\end{center}
\end{figure}

Figure~\ref{Fig:tri_phasediagram} shows the $H$-$K$ phase diagram for the triangular lattice case, which consists of six phases in addition to the fully-polarized state for $H \geq 2$. 
We describe the details of each phase in what follows. 
The real-space spin configuration and the spin and chirality structure factors for each phase are shown in Fig.~\ref{Fig:tri_eachphase}. 

\paragraph{triple-$Q'$ chiral stripe I (3$Q'$-I)}
This phase occupies a low-$H$ and low-$K$ region, appearing next to the double-$Q'$ chiral stripe at $H=0$. 
In this phase, the $z$ component of the spin structure factor has a single-$Q$ peak at $\mathbf{Q}_3$, while the $xy$ component exhibits two peaks at $\mathbf{Q}_1$ and $\mathbf{Q}_2$ with different intensities, as shown in Fig.~\ref{Fig:tri_eachphase}(a). 
Meanwhile, the chirality structure factor shows a single-$Q$ peak at $\mathbf{Q}_3$. 

\paragraph{triple-$Q$ $n_{\rm sk}=2$ skyrmion crystal (3$Q$ with $n_{\rm sk}=2$)}
This phase occupies a higher-$K$ region next to the triple-$Q'$ chiral stripe I state at low $H$,
evolving from the $H=0$ state with spin canting in the field direction. 
The $xy$ component of the spin structure factor exhibits two dominant peaks and a single subdominant peak, while the $z$ component shows a single dominant peak and two subdominant peaks, as shown in Fig.~\ref{Fig:tri_eachphase}(b). 
The amplitudes of the total spin structure factor at $\mathbf{Q}_1$, $\mathbf{Q}_2$, and $\mathbf{Q}_3$ are equivalent to each other. 

\paragraph{triple-$Q'$ chiral stripe II (3$Q'$-II)}
This phase occupies a slightly high-$H$ region of the triple-$Q$ $n_{\rm sk}=2$ skyrmion crystal, as shown in Fig.~\ref{Fig:tri_phasediagram}. 
The spin configuration shown in Fig.~\ref{Fig:tri_eachphase}(c) resembles that in the triple-$Q'$ chiral stripe I phase shown in Fig.~\ref{Fig:tri_eachphase}(a): it is characterized by the double-$Q$ modulation in the $xy$ component and the single-$Q$ modulation in the $z$ component of the spin structure factor. 
The difference is in the peak intensities of the $xy$ component: the two peaks have equal weights in the triple-$Q'$ chiral stripe II phase, while they are different in the triple-$Q'$ chiral stripe I phase. 
In the entire region of this phase, the amplitudes of the two peaks in the $xy$ component are almost the same as that of the single peak in the $z$ component, but the latter is slightly larger than the former. 

\paragraph{triple-$Q$ $n_{\rm sk}=1$ skyrmion crystal (3$Q$ with $n_{\rm sk}=1$)}
This phase is another type of crystallization of skyrmions, which has been found in chiral and frustrated magnets~\cite{Muhlbauer_2009skyrmion,yu2010real,nagaosa2013topological,Okubo_PhysRevLett.108.017206,leonov2015multiply,Lin_PhysRevB.93.064430,Hayami_PhysRevB.93.184413}. 
It occupies a large portion of the intermediate-$H$ region, for a slightly larger $H$ than the triple-$Q'$ chiral stripe II. 
Figure~\ref{Fig:tri_eachphase}(d) shows a typical real-space spin configuration in the skyrmion crystal. 
The skyrmion cores drawn by the blue regions form a triangular lattice with the lattice constant $4\pi/(\sqrt{3}|\mathbf{Q}_{\nu}|)$. 
The spin structure factors have six peaks with equal intensities, as shown in Fig.~\ref{Fig:tri_eachphase}(d), which indicates the presence of the $C_6$ symmetry. 
This phase also accompanies a uniform net scalar chirality, which is around half of that in the triple-$Q$ $n_{\rm sk}=2$ skyrmion crystal; this phase is characterized by the topological number $n_{\rm sk}=1$.  
The remarkable point of this skyrmion crystal is that it is stable even at $T=0$. This is in contrast to the similar state obtained for a localized spin model with isotropic Heisenberg interactions~\cite{Okubo_PhysRevLett.108.017206}, which exists only at nonzero temperature and is replaced with the single-$Q$ conical ordering at low temperature. 
It is also worth noting that the skyrmion crystal state obtained here shows the degeneracy with respect to the chiral symmetry as well as the rotational symmetry along the $z$ axis, in contrast to the absence in the Heisenberg model with the DM interaction~\cite{Lin_PhysRevB.93.064430}. 

\paragraph{triple-$Q'$ chiral stripe III (3$Q'$-III)}
This phase occupies the largest portion of the phase diagram in Fig.~\ref{Fig:tri_phasediagram}, appearing next to the triple-$Q$ $n_{\rm sk}=1$ skyrmion crystal upon increasing $H$. 
The spin configuration resembles that in the triple-$Q'$ chiral stripe II state, while the $xy$ component of the spin structure factor shows a small peak at $\mathbf{Q}_1 - \mathbf{Q}_2$, as shown in Fig.~\ref{Fig:tri_eachphase}(e). 
This phase is similar to that found in the Kondo lattice model at high field in Ref.~\onlinecite{ozawa2016zero}. 

\paragraph{triple-$Q'$ chiral stripe IV (3$Q'$-IV)}
This phase occupies a high-$H$ and high-$K$ region of the phase diagram in Fig.~\ref{Fig:tri_phasediagram}, next to the triple-$Q'$ chiral stripe III. 
In this phase, the spin configuration is also characterized by the triple-$Q$ modulation: 
the $xy$ component of the spin structure factor has two dominant peaks with additional four subdominant peaks, while the $z$ component shows four dominant peaks, as shown in Fig.~\ref{Fig:tri_eachphase}(f). 

\paragraph{Magnetization curve and net scalar chirality}
Figure~\ref{Fig:tri_mag}(a) shows the magnetization curve at several $K$. 
As in the square lattice case in Fig.~\ref{Fig:squ_mag}, the magnetization becomes nonzero for infinitesimal $H$ and continuously increases with increasing $H$ irrespective of $K$, which indicates continuous phase transition at $H=0$.  
With further increasing $H$, the magnetization jump appears between the triple-$Q$ $n_{\rm sk}=2$ skyrmion crystal and the triple-$Q'$ chiral stripe II. 
The similar magnetization jumps are found in phase boundaries between the triple-$Q$ $n_{\rm sk}=1$ skyrmion crystal and the triple-$Q'$ chiral stripe II, III. 
Thus, for $K=0.1$, two jumps in the magnetization are found at $H\simeq 0.5$ and $1.0$, whereas three jumps are found for $K=0.4$ and 0.8, at $H\simeq 0.15$, $0.30$, $0.95$ and $H\simeq 0.15$, $0.35$, $0.95$, respectively, as shown in Fig.~\ref{Fig:tri_mag}(a).  
Other phase transitions, e.g., between triple-$Q'$ chiral stripe and fully-polarized state are continuous. 

On the other hand, Fig.~\ref{Fig:tri_mag}(b) shows $H$ dependence of the net scalar chirality at several $K$. 
The net scalar chirality is given by $\chi^{\rm sc} = (1/N) \sum_{\mathbf{R} \in \mu=(u,d)} \chi^{\rm sc}_{\mathbf{R}}$, where $\mu =(u,d)$ represent upward and downward triangles, respectively.  
As shown in Fig.~\ref{Fig:tri_mag}(b), there are two phases which show the net scalar chirality, the triple-$Q$ $n_{sk}=2$ skyrmion crystal stabilized in the low $H$ region and the triple-$Q$ $n_{sk}=1$ skyrmion crystal in the intermediate $H$ region. 
In both phases, the scalar chirality is almost independent of $H$, while it changes discontinuously at their phase boundaries. 

\subsection{Comparison to the Kondo lattice model}

Let us compare our results with the previous studies in the Kondo lattice model. 
In Ref.~\onlinecite{ozawa2016zero}, the Kondo lattice model on the triangular lattice was studied in an applied magnetic field. 
There found the phase sequence from the triple-$Q$ $n_{\rm sk}=2$ skyrmion crystal, the triple-$Q$ $n_{\rm sk}=1$ skyrmion crystal, the triple-$Q'$ chiral stripe III (3$Q'$-III), and to the fully-polarized state with increasing $H$. 
Similar sequence is found in our result in the region $0.37 \lesssim K\lesssim 0.53$ in Fig.~\ref{Fig:tri_phasediagram}. 
The only difference is that the triple-$Q$ chiral stripe II appears between the triple-$Q'$ $n_{\rm sk}=2$ skyrmion crystal and the triple-$Q$ $n_{\rm sk}=1$ skyrmion crystal in our results. 
A possible reason for this discrepancy is due to the lack of the effect of the magnetic field on Green's function in our effective model, which gives rise to the spin dependent interaction via the Zeeman coupling in Eq.~(\ref{eq:zeeman}).  Among them, the third-order contributions with respect to $J$ arise in nonzero $H$, and may prefer the triple-$Q$ skyrmion crystals to the triple-$Q$ chiral stripe; see Appendix~\ref{sec:Third-order effective magnetic interactions}. 
Another possibility is the charge degree of freedom in the Kondo lattice model.
In our effective spin model, it is implicitly assumed that the electron density is unchanged while changing the magnetic field, while the results in the Kondo lattice model were calculated for a fixed chemical potential~\cite{ozawa2016zero}. Such field dependence of the electron density may result in the discrepancy. 

Meanwhile, for the square lattice case, the authors have found the phase transition from the double-$Q'$ chiral stripe to the double-$Q$ state in the Kondo lattice model with increasing $H$ in Ref.~\onlinecite{Ozawa_chiralstripe}.  This corresponds to the region for $ 0.27 \lesssim K$ in our effective spin model. 
This agreement supports the validity of our effective model.

Although the detailed comparison is left for future study, we stress that our effective model reproduces well the exotic magnetic phases found in the Kondo lattice model. 
The effective spin model will be useful to explore further exotic phases by smaller computational efforts compared to those for itinerant electron models. 

\section{Summary and Concluding Remarks}
\label{sec:Discussion}

To summarize, we have constructed an effective spin model for describing magnetic instabilities in itinerant magnets. 
Taking one of the fundamental models for itinerant magnets, the Kondo lattice model, and carefully examining the spin scattering processes by the perturbation in terms of the spin-charge coupling, we proposed a simple  model including the bilinear and biquadratic interactions with particular wave numbers dictated by the Fermi surface. 
The bilinear interaction has the same form as the RKKY interaction, and prefers a helical magnetic order specified by a single wave number. 
On the other hand, the biquadratic interaction, which is deduced from the dominant contributions in the higher-order perturbations, causes an instability toward noncollinear or noncoplanar ordering specified by multiple wave numbers. 
We have tested the validity of the effective model by calculating the ground-state phase diagrams by Monte Carlo simulation and comparing the results with those for the Kondo lattice model. 
The comparison on square and triangular lattices shows a good agreement with the previous findings in the Kondo lattice model~\cite{Ozawa_doi:10.7566/JPSJ.85.103703,ozawa2016zero}: we have demonstrated that our effective spin model reproduces the noncoplanar double-$Q$ instability in both square and triangular lattice cases and the triple-$Q$ skyrmion with topological number of two in the triangular lattice case. 
The good agreement indicates that our effective model captures the underlying mechanism of the magnetic instabilities toward noncollinear and noncoplanar orderings in itinerant magnets: the key ingredient is the competition between bilinear and biquadratic interactions in momentum space. 

We have also extended our study by introducing the external magnetic field to our effective spin model. 
The simplicity of the model allows us to systematically investigate the magnetic phase diagram by much smaller computational costs compared to the original Kondo lattice model. 
In addition to the phases already found in the previous studies for the Kondo lattice model~\cite{ozawa2016zero}, we found several new phases in an applied magnetic field. 
The results demonstrate the efficiency of our model for further exploration of exotic magnetic phases in itinerant magnets. 

Our effective spin model is constructed on the basis of a fundamental property of the spin-charge system: multiple peaks in the bare susceptibility whose wave numbers are related with the lattice symmetry. 
As this is commonly seen in all the centrosymmetric lattices in two and three dimensions, we believe that our model is applicable to a wide class of itinerant magnets. 
In particular, the instability toward noncollinear and noncoplanar orderings induced by the effective biquadratic interaction will be a universal feature irrespective of the details of the system. 
Also, we note that our effective spin model can be easily extended to more complicated situations, by including, e.g., anisotropic interactions, single-ion anisotropy, and dipole-dipole interactions. 
Such extensions will be helpful to investigate and revisit unconventional magnetic phase diagrams from the viewpoint beyond the RKKY mechanism. 

Finally, let us comment on candidate materials, whose magnetic properties might be described by our spin model. 
One of the candidates is a scandium thiospinel MnSc$_2$S$_4$, which has been recently identified to show a triple-$Q$ vortex crystal under an external magnetic field~\cite{Gao2016Spiral}. 
Another candidate is the strontium iron perovskite oxide SrFeO$_3$, which shows several phases depending on magnetic field and temperature~\cite{Ishiwata_PhysRevB.84.054427}. 
Interestingly, the transport measurements imply that the phases include multiple-$Q$ states in spite of the negligibly small contribution of the spin-orbit coupling. 
Our model will make a significant contribution to understand these exotic magnetisms. 
The detailed comparison will be left for future study. 

\begin{acknowledgments}
R.O. is supported by the Japan Society for the Promotion of Science through a research fellowship for young scientists and the Program for Leading Graduate Schools (ALPS). 
This research was supported by KAKENHI (No.~16H06590), the Strategic Programs for Innovative Research (SPIRE), MEXT, and the Computational Materials Science Initiative (CMSI), Japan. 
\end{acknowledgments}

\appendix
\section{Third-order effective magnetic interactions in a magnetic field}
\label{sec:Third-order effective magnetic interactions}

\begin{figure}[htb!]
\begin{center}
\includegraphics[width=1.0 \hsize]{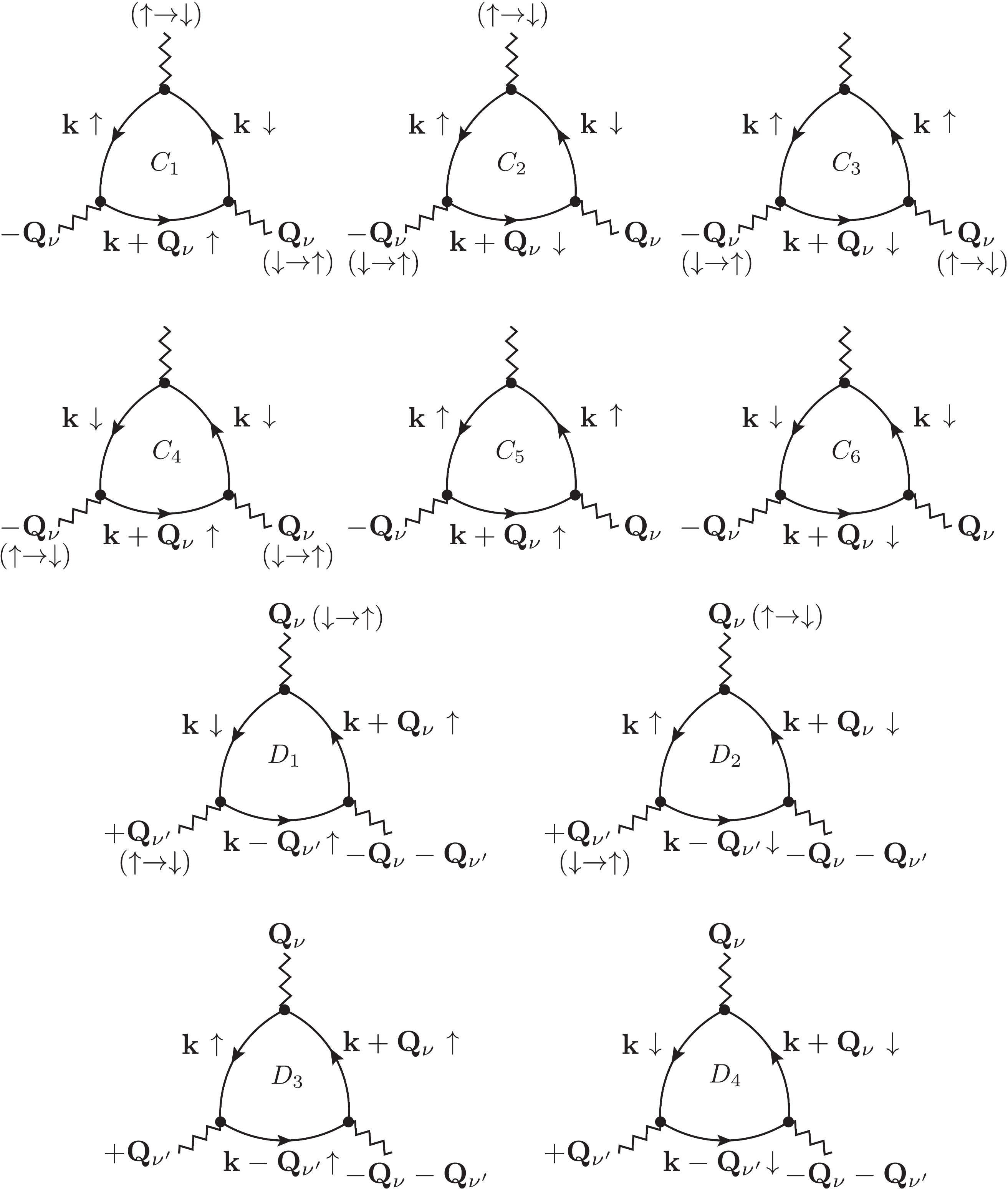} 
\caption{
\label{Fig:diagram_3rd}
Feynman diagrams for the coefficients $C_1$, $C_2$, $C_3$, $C_4$, $C_5$, $C_6$, $D_1$, $D_2$, $D_3$, and $D_4$ in the third-order contributions. See Eqs.~(\ref{eq:C1})-(\ref{eq:D4}).
}
\end{center}
\end{figure}

In this appendix, we show the expression of the third-order magnetic interactions, which appears under the external magnetic field. 
By substituting $\mathbf{q}=\mathbf{Q}_1$, $\mathbf{Q}_2$, and $\mathbf{Q}_3$ in the expansion of the free energy in Eq.~(\ref{eq:freeenergy_expand}), we obtain the third-order contributions, which are given by 
\begin{widetext}
\begin{align}
F^{(3)}&=F^{(3)}_1+F^{(3)}_2+F^{(3)}_3+F^{(3)}_4+F^{(3)}_5, \\
F^{(3)}_1 &=-2 \frac{J^3}{\sqrt{N}} \sum_{\nu}
(C_1-C_2)\left[ S_{\mathbf{Q}_\nu}^z (S_{\mathbf{0}}^x S^x_{-\mathbf{Q}_\nu}+S_{\mathbf{0}}^y S^y_{-\mathbf{Q}_\nu}) + {\rm H.c.} \right], \\
F^{(3)}_2&=-2 \frac{J^3}{\sqrt{N}} \sum_{\nu}
(C_3-C_4)\left[ S_{0}^z (S_{\mathbf{Q}_\nu}^x S^x_{-\mathbf{Q}_\nu}+S_{\mathbf{Q}_\nu}^y S^y_{-\mathbf{Q}_\nu}) + {\rm H.c.} \right], \\
F^{(3)}_3&=-2 \frac{J^3}{\sqrt{N}} \sum_{\nu}
(C_5-C_6) S_{0}^z S_{\mathbf{Q}_\nu}^z S^z_{-\mathbf{Q}_\nu}, \\
F^{(3)}_4&=-2 \frac{J^3}{\sqrt{N}}
(D_1-D_2) \left[ S_{\mathbf{Q}_1}^z (S_{\mathbf{Q}_2}^x S^x_{\mathbf{Q}_3}+S_{\mathbf{Q}_2}^y S^y_{\mathbf{Q}_3}) + {\rm H.c.} \right] \nonumber \\ 
&+(\mathbf{Q}_1 \to \mathbf{Q}_2, \mathbf{Q}_2 \to \mathbf{Q}_3, \mathbf{Q}_3 \to \mathbf{Q}_1)+(\mathbf{Q}_1 \to \mathbf{Q}_3, \mathbf{Q}_2 \to \mathbf{Q}_1, \mathbf{Q}_3 \to \mathbf{Q}_2), \\
F^{(3)}_5&=-2 \frac{J^3}{\sqrt{N}}
(D_3-D_4) \left[S_{\mathbf{Q}_1}^z S_{\mathbf{Q}_2}^z S^z_{-\mathbf{Q}_3}+ {\rm H.c.}\right], 
\end{align}
\end{widetext}
where H.c. represent the Hermitian conjugate. The coefficients are represented by 
\begin{align}
\label{eq:C1}
C_1 &=\frac{T}{N}\sum_{\mathbf{k}, \omega_p} G_{\mathbf{k}\uparrow}G_{\mathbf{k}\downarrow}G_{\mathbf{k}+\mathbf{Q}_\nu \uparrow}, \\
C_2 &=\frac{T}{N}\sum_{\mathbf{k}, \omega_p} G_{\mathbf{k}\uparrow}G_{\mathbf{k}\downarrow}G_{\mathbf{k}+\mathbf{Q}_\nu \downarrow}, \\
C_3 &=\frac{T}{N}\sum_{\mathbf{k}, \omega_p} G^2_{\mathbf{k}\uparrow}G_{\mathbf{k}+\mathbf{Q}_\nu \downarrow}, \\
C_4 &= \frac{T}{N}\sum_{\mathbf{k}, \omega_p}G^2_{\mathbf{k}\downarrow}G_{\mathbf{k}+\mathbf{Q}_\nu \uparrow}, \\
C_5 &= \frac{T}{N}\sum_{\mathbf{k}, \omega_p}G^2_{\mathbf{k}\uparrow}G_{\mathbf{k}+\mathbf{Q}_\nu \uparrow}, \\
C_6 &= \frac{T}{N}\sum_{\mathbf{k}, \omega_p}G^2_{\mathbf{k}\downarrow}G_{\mathbf{k}+\mathbf{Q}_\nu \downarrow}, \\
D_1 &= \frac{T}{N}\sum_{\mathbf{k}, \omega_p}G_{\mathbf{k}\downarrow}G_{\mathbf{k}+\mathbf{Q}_\nu \uparrow}G_{\mathbf{k}-\mathbf{Q}_{\nu'} \uparrow}, \\
D_2 &= \frac{T}{N}\sum_{\mathbf{k}, \omega_p}G_{\mathbf{k}\uparrow}G_{\mathbf{k}+\mathbf{Q}_\nu \downarrow}G_{\mathbf{k}-\mathbf{Q}_{\nu'} \downarrow}, \\
D_3 &=\frac{T}{N}\sum_{\mathbf{k}, \omega_p} G_{\mathbf{k}\uparrow}G_{\mathbf{k}+\mathbf{Q}_\nu \uparrow}G_{\mathbf{k}-\mathbf{Q}_{\nu'} \uparrow}, \\
\label{eq:D4}
D_4 &= \frac{T}{N}\sum_{\mathbf{k}, \omega_p}G_{\mathbf{k}\downarrow}G_{\mathbf{k}+\mathbf{Q}_\nu \downarrow}G_{\mathbf{k}-\mathbf{Q}_{\nu'} \downarrow},
\end{align} 
where $C_1$ to $C_6$ represent the scattering processes by a single-$Q$ wave number, while $D_1$ to $D_4$ the scattering processes by multiple-$Q$ wave numbers. 
We may need to take into account these terms in the effective spin model in nonzero field, which is left for future study. 


\nocite{apsrev41Control}
\bibliographystyle{my-apsrev4-1}
\bibliography{my-refcontrol,ref}

\end{document}